\documentclass[ApJL, twocolumn]{aastex631}

\usepackage{graphicx}
\usepackage{amsmath}
\usepackage{color}
\usepackage[normalem]{ulem}
\usepackage{units}
\usepackage{soul}
\usepackage{algorithm}
\usepackage{algpseudocode}

\newcommand{\SPA}{School of Physics and Astronomy, Monash University, Vic 3800, Australia}
\newcommand{\OzGravMonash}{OzGrav: The ARC Centre of Excellence for Gravitational Wave Discovery, Clayton VIC 3800, Australia}

\newcommand{\Caltech}{Department of Physics, California Institute of Technology, Pasadena, California 91125, USA}

\newcommand{\pistroke}{
  \text{\protect\ooalign{\hidewidth\raisebox{-0.2ex}{--}\hidewidth\cr$\pi$\cr}}
}
\newcommand{\Lstroke}{%
  \text{\protect\ooalign{\hidewidth\raisebox{0.2ex}{--}\hidewidth\cr$\mathcal{L}$\cr}}%
}

\begin{document}

\title{Trends in the Population of Binary Black Holes Following the Fourth \\
Gravitational-Wave Transient Catalog: a Data-Driven Analysis}

\author{Nir Guttman}
\email{nir.guttman@monash.edu}
\affiliation{\SPA}
\affiliation{\OzGravMonash}

\author{Ethan Payne}
\affiliation{\Caltech}

\author{Paul D. Lasky}
\affiliation{\SPA}
\affiliation{\OzGravMonash}

\author{Eric Thrane}
\affiliation{\SPA}
\affiliation{\OzGravMonash}

\begin{abstract}
Current population models of binary black hole distributions are difficult to interpret because standard population inferences hinge on modeling choices, which can mask or mimic real structure.  
The maximum population likelihood ``$\pistroke$ formalism'' provides a means to investigate and interpret features in the distribution of binary black holes using only data---without specifying a population model.  
It tells us if features inferred from current population models are truly present in the data or if they arise from model misspecification.
It also provides guidance for developing new models by highlighting previously unnoticed features.
In this study, we utilize the $\pistroke$ formalism to examine the binary black hole population in the LIGO--Virgo--KAGRA (LVK) fourth Gravitational-Wave Transient Catalog (GWTC-4). 
Our analysis supports the existence of a gap around $45\,M_\odot$ in the secondary black hole mass distribution and identifies a widening in the distribution of the effective inspiral spin parameter $\chi_\text{eff}$ near this mass as recently reported by \citet{Tong_mass_gap}. 
Similar to earlier studies, we find support for an anti-correlation between $\chi_\text{eff}$ and mass ratio.
However, we argue that this may be a spurious correlation arising from misspecification of the joint distribution of black hole masses. 
Furthermore, we identify support for dimensionless black hole spin magnitudes at approximately $\chi \approx 0.2$ and $\chi\approx0.7$.
The data support the existence of a correlation between the spin magnitudes $\chi_1$ and $\chi_2$, though subsequent study is required to determine if this feature is statistically significant. 
The accompanying data release includes $\pistroke$ samples, which can be used to compare theoretical predictions to LVK data and to assess assumptions in parameterised models.
\end{abstract}

\section{Introduction}
Hierarchical Bayesian inference is a standard approach for inferring the properties of populations from ensembles of uncertain observations. In this framework, one assumes a family of population distributions parameterized by hyperparameters, which are inferred from multiple measurements. This methodology consistently accounts for measurement uncertainty and selection effects, and is widely employed in gravitational-wave astrophysics and other fields~\citep[e.g.,][]{intro,Vitale2020,o3a_pop,gwtc-3_pop,GWTC-4_pop}.

A fundamental limitation of hierarchical inference is its dependence on model assumptions: the population distribution must be specified before fitting. Although parametric models are designed to capture key population features, they include assumptions that may not be consistent with the true distributions in nature. Consequently, important features of the true population may be distorted or entirely overlooked, depending on the model’s flexibility and structure. Such \textit{model misspecification} becomes increasingly problematic as the quantity and precision of available data increase~\citep{wmf}.
The problem is somewhat ameliorated with data-driven models, which use a large number of parameters to provide the flexibility to mimic a wide range of distributions~\citep[e.g.,][]{Tiwari_2021,Tiwari_2022,B_spline,PhysRevX.14.021005,pixel_pop}.
However, even highly-parameterised, data-driven models are influenced by the choice of hyperparameter prior, which can obfuscate features present in the data.

The $\pistroke$ formalism (pronounced ``pi stroke'') is a framework for model exploration and data visualisation, which eschews models entirely~\citep{pi_stroke_orig}. 
In this approach, one finds the maximum population likelihood solution---the distribution that maximizes the population likelihood over all possible population models.
For example, one can find the distribution of primary black hole mass $m_1$ that provides the maximum likelihood solution compared to all possible distributions of $m_1$.
The maximum population likelihood distribution is denoted $\pistroke$.

The $\pistroke$ solution is a sum of delta functions, with heights and positions in parameter space that maximize the population likelihood. 
The $\pistroke$ distribution is derived entirely from the data and provides a means of visualising the data, free from explicit model constraints. 
It has been conjectured by~\citet{pi_stroke_orig} that, in the limit that the number of observations goes to infinity, $\pistroke$ reproduces the true distribution in nature.

Features inferred by population models that do not appear in $\pistroke$ can be evidence of model misspecification and/or extrapolation.
Features that do appear in $\pistroke$ are not necessarily statistically significant (or physical), but they are at least supported by the data.
Bayesian inference is still required to determine which features are significant.

Gravitational-wave astronomy offers a compelling application for this methodology. Population inference is central to understanding the formation and evolution of compact-object binaries. 
The observed population of binary black hole mergers is now sufficiently large to support meaningful population-level analyses, due to the improved sensitivity of the LVK network~\citep{detector_PhysRevD.102.062003,detector_PhysRevX.13.041021,detector_doi:10.1126/science.ado8069,detector_PhysRevD.111.062002,detector_Soni_2025}. 
Recent studies~\citep{o3a_pop,gwtc-3_pop,GWTC-4_pop} have identified complex structures in distributions of black hole mass, spin, and redshift.

\citet{pi_stroke_orig} applied the $\pistroke$ formalism to data from the third Gravitational-Wave Transient Catalog~\citep[GWTC-3;][]{GWTC-3}.
The $\pistroke$ distributions in~\citet{pi_stroke_orig} mostly supported conclusions from the LIGO--Virgo--KAGRA analysis~\citep{gwtc-3_pop}. 
Both approaches found a primary-mass distribution that falls off steeply yet exhibits pronounced bumps and dips, rather than a single smooth power law.
Neither evidence provided suggested support for a hard upper mass gap above $\approx 60 M_\odot$. 
The effective inspiral spin parameter $\chi_\text{eff}$---a mass-weighted projection of the component spins onto the orbital angular momentum---showed essentially no support for negative values in the $\pistroke$ analysis, suggesting that the modest negative tail found by \citet{gwtc-3_pop} was model-dependent. 
Both analyses indicated that the merger-rate density grows with redshift, consistent with a simple power-law evolution.

In this work, we apply the $\pistroke$ formalism to the new GWTC-4 catalog of binary black hole mergers~\citep{GWTC-4}. 
We enhance the numerical implementation to accommodate the larger dataset and investigate the structures preferred by the data. 
Our goal is to provide a data-driven perspective on binary black hole population properties, reveal previously overlooked features, assess the support for existing models, and inform future model development efforts.

The remainder of this paper is organized as follows.
Section~\ref{sec:method} summarizes the $\pistroke$ formalism.
Section~\ref{sec:results} presents the $\pistroke$ results, and Section~\ref{sec:discussion} discusses the implications.
Our concluding remarks are presented in Section~\ref{sec:conclusions}.

\section{Methodology}\label{sec:method}
\subsection{Background}
Throughout this work we use the standard notation in gravitational-wave science: $m_1\ge m_2$ are the component masses, $q=m_2/m_1\le1$ is the mass ratio, $z$ is redshift, $\chi_1,\chi_2$ are dimensionless spin magnitudes, and $\cos\theta_{1},\cos\theta_2$ are the tilt angles between each spin vector and the binary orbital angular momentum.
The quantity
\begin{align}
\chi_{\mathrm{eff}}=(\chi_1\cos\theta_1+q\,\chi_2\cos\theta_2)/(1+q),
\end{align}
is the effective inspiral spin parameter, which is an approximate constant of motion~\citep{Damour_2001}.
Meanwhile, $\chi_{\mathrm{p}}$ is the effective precession spin parameter~\citep{chi_p}:
\begin{align}
\chi_{\mathrm{p}}=\max\left(\chi_1 \sin\theta_1,\ \left(\frac{ 4q + 3}{4 + 3q}\right) q\chi_2 \sin\theta_2\right).
\end{align}

Hierarchical Bayesian inference provides a robust framework for inferring population-level properties from multiple uncertain observations. The population likelihood is defined as the joint likelihood of the dataset given a population distribution, integrated over the parameters of each individual observation:
\begin{equation}
\mathcal{L}(d|\Lambda, M) = \prod_{i=1}^{N}\frac{1}{\xi(\Lambda)}\int d\theta_i\,\mathcal{L}(d_i|\theta_i)\,\pi(\theta_i|\Lambda, M),
\label{eq:pop_likelihood}
\end{equation}
where \(\mathcal{L}(d_i|\theta_i)\) is the likelihood of observation of data \(d_i\) given binary parameters \(\theta_i\), and \(\pi(\theta_i|\Lambda, M)\) is the conditional prior defining the population model $M$, describing the distribution of binary parameters given hyper-parameters \(\Lambda\). 
Meanwhile, \(N\) is the total number of observations. The normalization factor \(\xi(\Lambda)\) accounts for observational selection effects:
\begin{equation}
\xi(\Lambda) = \int d\theta\, P_{\text{det}}(\theta)\pi(\theta|\Lambda, M) .
\label{eq:selection_effect}
\end{equation}
Here, $P_{\text{det}}(\theta)$ denotes the probability that an event with astrophysical parameters $\theta$ is detected. For further details on hierarchical Bayesian inference and selection effects in gravitational-wave astronomy, see~\citet{selction_effects,Thrane_Talbot_2019,Vitale2020}.

Both a feature and a limitation of hierarchical Bayesian inference is its dependence on a model, \(M\). 
Bayesian inference can tell us which of two models is preferred, but in its standard formulation, it does not tell us if our best model is adequate~\citep[e.g.,][]{wmf}.
The $\pistroke$ formalism can address potential issues arising from model misspecification~\citep{pi_stroke_orig}.
This formalism identifies the distribution $\pistroke$ that maximizes the population likelihood (Eq.~\eqref{eq:pop_likelihood}) across all possible distributions:
\begin{equation}
\text{\Lstroke} \equiv \max_{\pi}\,\mathcal{L}(d|\pi).
\label{eq:l_stroke}
\end{equation}

The likelihood–maximizing distribution (i.e., the argmax in Eq.~\eqref{eq:l_stroke}) admits a discrete representation as a weighted sum of $n$ delta functions:
\begin{equation}
\text{\pistroke}(\theta) = \sum_{k=1}^{n} w_k \,\delta(\theta - \theta_k),
\label{eq:pi_stroke}
\end{equation}
where the weights $w_k$ satisfy $\sum_{k=1}^{n} w_k = 1$, and $\theta_k$ represent parameter locations maximizing the likelihood. 
The number of delta functions $n$ is always less than or equal to the number of measurements $N$.
This discrete representation, guaranteed by Carathéodory's theorem~\citep{Carathodory1911berDV}, provides an upper bound on the population likelihood $\Lstroke$.
The set of delta functions explicitly represents regions in parameter space best supported by the observed data.
By comparing the maximum likelihood value from a parametric model to $\Lstroke$, one can determine if the model captures the salient features present in the data.
Substituting Eq.~\eqref{eq:pi_stroke} into Eq.~\eqref{eq:pop_likelihood} and Eq.~\eqref{eq:selection_effect} yields a simplified expression:
\begin{equation}
\text{\Lstroke} = \prod_{i=1}^{N}\frac{1}{\xi(\text{\st{M}})}\sum_{k=1}^{n} w_k\,\mathcal{L}(d_i|\theta_k),
\label{eq:l_stroke_final}
\end{equation}
where
\begin{equation}
\xi(\text{\st{M}}) = \sum_{k=1}^{n} w_k\,P_{\text{det}}(\theta_k) ,
\label{eq:xi_stroke}
\end{equation}
is the normalisation factor for the distribution $\pistroke$.
The $\pistroke$ distribution defines what we call a pseudo-model (denoted $\text{\st{M}}$).
It is not a true model because it is determined by the data whereas a true model must be defined before considering the data.

The ratio of the number \(n\) of delta functions to the total number of observations \(N\) defines the informativeness:
\begin{equation}
\mathcal{I} \equiv \frac{n}{N}.
\label{eq:informativeness}
\end{equation}
This metric quantifies the resolvability of different parameter distributions.

\subsection{Data and selection effects}
\label{sec:selection_effects}
We utilize data from GWTC-4~\citep{GWTC-4}, selecting events with a false alarm rate less than $1\,\text{yr}^{-1}$. The resulting dataset comprises 153 binary black hole mergers. Throughout this analysis, we make use of the preferred posterior samples from GWTC-4~\citep{gwtc2_data,gwtc3_data,gwtc4_data}.

To account for selection effects in our pseudo-model \st{M}, we start from Eq.~\eqref{eq:selection_effect}, and decompose the distribution \(\pi(\theta|\Lambda,M)\) into three distinct terms:
\begin{align}
\pi(\theta|\hat\Lambda,\text{\st{M}}) = \pi(\eta | \hat\Lambda) \, \pi(\zeta | \text{\o}) \sum_k w_k \, \delta(m - m_k).
\end{align}
The first term \(\pi(\eta | \hat\Lambda)\) represents the population model for parameters not directly analyzed in the $\pistroke$ framework, with \(\hat{\Lambda}\) the hyper-parameters chosen to maximize the population likelihood. 
For example, if we make $\pistroke(m_1, m_2)$, we need to make some assumption about the distributions of black hole spins.
The second term \(\pi(\zeta|\text{\o})\) denotes the distribution for nuisance parameters with trivial priors, e.g., the gravitational-wave polarization angle. The third term consists of delta functions localized at parameters of interest, which are denoted \(m_k\).

Substituting this decomposition into Eq.~\eqref{eq:selection_effect} yields:
\begin{align}
    P_\text{det}(\text{\st{M}}) &= \int d\zeta \int d\eta \sum_k w_k \, P_\text{det}(m_k, \eta, \zeta) \, \pi(\eta | \hat\Lambda) \, \pi(\zeta | \text{\o}) \\ \nonumber
    &= \int d\eta \sum_k w_k \, P_\text{det}(m_k, \eta) \, \pi(\eta | \hat\Lambda),
\end{align}
where 
\begin{align}
    P_\text{det}(m_k, \eta) \equiv \int d\zeta \, P_\text{det}(m_k, \eta, \zeta) \, \pi(\zeta | \text{\o}) .
\end{align}

In order to estimate $P_\text{det}$, we use the official LIGO–Virgo--KAGRA list of found injections for GWTC-4~\citep{essick_injection}. The found injection samples provides a faithful, catalog-specific estimate of the selection function~\citep{GWTC-4_pop}.
The detection probability $P_\text{det}(m,\eta)$ can be written in terms of the distribution of found injections $f_\text{det}(m,\eta)$ and the distribution of injections $\pi(m,\eta|\text{\o})$:\footnote{By Bayes' rule,
$p(\mathrm{det}\mid m,\eta)=p(m,\eta\mid \mathrm{det})\,\pi(\mathrm{det})/\pi(m,\eta)$.
Identifying $f_{\mathrm{det}}(m,\eta)\equiv \pi(m,\eta\mid \mathrm{det})$,
$\pi(m,\eta\mid \text{\o})\equiv\pi(m,\eta)$, and $n_{\mathrm d}/N\equiv \pi(\mathrm{det})$
yields Eq.~\eqref{eq:p_det_def}.} 
\begin{align}
\label{eq:p_det_def}
    P_\text{det}(m, \eta) = \frac{f_\text{det}(m, \eta)}{\pi(m, \eta|\text{\o})} \frac{n_{\text{d}}}{N} .
\end{align}
Here, \(n_{\text{d}}\) is the number of found injections and \(N\) is the total number of injections. 
This result implies:
\begin{align}
    P_\text{det}(\text{\st{M}}) &= \frac{n_{\text{d}}}{N} \int d\eta \sum_k w_k \frac{f_\text{det}(m_k,\eta)}{\pi(m_k,\eta|\text{\o})} \pi(\eta | \hat\Lambda) \nonumber \\
    &= \frac{n_{\text{d}}}{N}\sum_{j\in\hat{\Lambda}, k} w_k \frac{f_\text{det}(m_k, \eta_j)}{\pi(m_k, \eta_j | \text{\o})},
\end{align}
with the summation over \(j\) performed using samples drawn from the distribution \(\pi(\eta|\hat\Lambda)\).

The empirical distribution \(f_\text{det}(\theta, \eta)\) can be effectively approximated using density estimation methods such as kernel density estimators~\citep{scott1992multivariate}. 
However, in this study, we adopt normalizing flows implemented via \textsc{FlowJax}~\citep{flowjax-doc} due to their flexibility and efficiency. Specific details about the flow architecture are provided in Appendix~\ref{appx:nf_method}. This approach to modeling selection effects in gravitational-wave astronomy builds on ideas from~\citet{p_det_colm}, where Gaussian mixture models were employed to approximate \(f_\text{det}\).
A similar approach proposed by~\citet{callister_p_det} models $P_{\text{det}}(\theta)$ directly using a neural network emulator;
for additional frameworks, see~\citet{p_det_NN,p_det_physics}.

\subsection{Marginal likelihoods}
\label{sec:marg_like_subset}
The $\pistroke$ calculation makes use of the marginal likelihood of data given a subset of parameters \(m\) marginalising over the remaining nuisance
parameters \(\eta\).
We weight LVK samples according to the weight
\begin{equation}
\label{eq:marg_weights}
  w_{j}
    \;=\;
    \frac{U(m_j)\,\pi(\eta_{j} \mid \hat{\Lambda})}
         {\pi(m_{j},\eta_{j} \mid \text{\o})} .
\end{equation}
The denominator undoes the fiducial prior used to generate the LVK samples $\pi(m_{j},\eta_{j} \mid \text{\o})$.
The numerator applies the population model to the nuisance parameters $\eta$ and a uniform prior to the parameter of interest $m$.
We then fit these weighted samples with a kernel density estimator if $m$ is a single parameter and using a normalising flow if $m$ consists of more than one parameter (see Appendix~\ref{appx:nf_method} for additional details).
The fit, which is a representation of the marginal posterior, is denoted $p_{\cal D}(m | d)$.

The associated Bayesian evidence for these new weights is given by
\begin{align}
    {\cal Z}_{\text{\o}} \sum_j w_j .
\end{align}
Here, ${\cal Z}_{\text{\o}}$ is the fiducial evidence from LVK parameter estimation.
The marginal likelihood is then 
\begin{align}
    {\cal L}(d | m) = & 
    \frac{p_{\cal D}(m | d)}{U(m)} \,
    {\cal Z}_{\text{\o}} \sum_j w_j \\
    \propto & p_{\cal D}(m | d).
\end{align}
Since we are interested only in finding the distribution of $m$ that maximises the population likelihood, we can ignore terms that do not depend on the distribution of $m$ as they are effectively constants.

For almost all events the effective sampling efficiency,
\begin{equation}
  \epsilon \;=\;
  \frac{\bigl(\sum_{j} w_{j}\bigr)^{2}}
       {\sum_{j} w_{j}^{2}},
\end{equation}
exceeds the \(1\%\) threshold recommended by~\citet{Ethan_eff}.  Only three events, all in the redshift analysis, dip below \(1\%\) but remain above \(0.5\%\). 

This weighting procedure requires us to assume a population model for the nuisance parameters. Different assumptions can, in principle, lead to different $\pistroke$ results. However, their influence should propagate primarily through correlations between the nuisance parameters and the parameters of interest. Consequently, as long as the adopted population models are qualitatively similar, we expect the resulting inferences to be robust.

To avoid circularity when analyzing GWTC-4 data and to enable a fair comparison with its parametric results, we adopt the GWTC-3~\citep{gwtc-3_pop} maximum-likelihood population model as an informed baseline. Details of the population model $\pi(\eta \mid \hat{\Lambda})$ used in the marginalization are provided in Appendix~\ref{appx:pop_model}.

Additional implementation details for the effective spin parameters 
\(\chi_{\mathrm{eff}}\), and \(\chi_{\mathrm{p}}\) are provided in
Appendix~\ref{appx:chi_eff_likelihood_weights}.

\subsection{Optimization procedure}
\label{sec:optimisation}
Having defined the components of $\Lstroke$ (Eq.~\eqref{eq:l_stroke_final}),
our next task is to identify the delta-function locations and weights that maximize $\Lstroke$. 
This is a high-dimensional optimization problem.
For one-dimensional $\pistroke$ calculations with $n$ delta functions, there are $2n-1$ delta-function parameters to scan: $n-1$ weights and $n$ locations.
For two-dimensional calculations, there are $3n-1$ such parameters.
We employ a multi-stage optimization scheme that combines warm-start initialization, gradient descent, and iterative pruning.
This procedure is designed to come close to the true global maximum while producing highly interpretable plots.
To this end, we merge and discard delta functions if we can do so without an appreciable loss of likelihood; this helps produce easy-to-understand figures.
An annotated flowchart of the full optimization workflow is provided in Appendix~\ref{app:pistroke_algorithm}.

\paragraph{Stage\,1: warm-start initialization.}
For each event we assign a delta function.
The location of the delta function is chosen randomly from an approximation of the marginal likelihood ${\cal L}(d_i | m)$ to weight the probability of different $m$ values.
The stochasticity of this step is designed to assist with convergence by better exploring the likelihood surface.
Each delta function is assigned an equal initial
weight.  
From this random starting point we perform 30 gradient-descent
steps to refine the parameters.  The entire procedure is repeated
40 times, and the trial that reaches the highest likelihood value
provides the initial point for Stage 2.

\paragraph{Stage\,2: gradient descent.}
Starting from the selected warm start, we run a long gradient-descent
chain with an exponentially decaying learning rate; early stopping
terminates the run when there is no improvement in log-likelihood for
1000 consecutive steps.  The optimisation is carried out in an unbounded
parameter space, with a smooth \(\tanh\) transformation that enforces
physical bounds on the mass, spin, and redshift parameters. As a result, unphysical regions are excluded from the optimization domain.

\paragraph{Stage\,3: pruning by merging and discarding.}
Once gradient descent converges, we examine the set of components in two stages:
\begin{enumerate}
  \item \emph{Merging.} 
  When two delta functions get close enough together, they are determined to be a single delta function and merged.
  Two components separated by less than 
  $5\%$ of the hyper-diagonal of the optimisation box\footnote{The $5\%$ value is introduced to reduce computational cost, and its value was determined by trial and error.} are merged if the absolute change in log-likelihood is \(\le 0.05/\sqrt{n}\), where $n$ is the number of delta-functions. The \(\sqrt{n}\) factor guards against drift in the log-likelihood, ensuring that many individually small merges do not accumulate into a large overall change.

  \item \emph{Discarding.} 
  If a delta function can be removed without significantly affecting the population likelihood, it is discarded as unnecessary.
  A component is removed and its weight redistributed among the remaining delta functions if the resulting log-likelihood decreases by at most \(0.05/\sqrt{n}\).  
\end{enumerate}

Merging nearby and discarding low-impact delta functions yields a simpler representation that retains only data-supported features, highlights the salient structure, improves readability, and reduces the risk of over-interpretation.
\paragraph{Stage\,4: iterative refinement.}
After pruning, we rerun stage 2, the long gradient-descent phase followed by stage 3, the
merge/discard step. The cycle terminates when two successive rounds
return the same number of components and no further improvement in
log-likelihood is achieved.

\paragraph{Stage\,5: importance ordering.}
For the last stage, we rank the surviving components by their standalone
log-likelihood contributions.  Components are re-introduced one at a
time, in descending order of importance, to test whether any can still
be removed with negligible loss in global likelihood; none of the models
presented in this work requires additional pruning after this step.

\paragraph{Stage\,6: rinse and repeat.}
Because the optimization landscape can contain multiple local maxima, we repeat the entire optimization pipeline \(\mathcal{O}(10^{2})\) times with different random seeds and retain the trial that achieves the highest log-likelihood.  In most cases the resulting log-likelihood values form a clear plateau, indicating convergence to the global maximum.
For the few cases without a plateau, we increase the number of restarts to 500 and verify that the highest-likelihood solutions reproduce the same qualitative $\pistroke$ structure.  
This six-stage procedure reliably converges to representative $\pistroke$ plots.

\section{Results}
\label{sec:results}
In this section we present the $\pistroke$ distributions inferred from the \textsc{GWTC–4} catalog.  We organize the discussion around three principal population properties: (i) masses, (ii) spins, and (iii) redshift evolution.
We restrict attention here to one- and two-dimensional $\pistroke$ analyses.  
Three-dimensional $\pistroke$ studies remain a goal for future study.

In each one-dimensional plot, black markers and vertical lines denote the recovered delta-function components, with their heights proportional to the corresponding weights. 
The normalisation is arbitrary, but tuned to aid comparison with parametric models.
The red bars show a histogram of these $\pistroke$ samples.  
The grey band indicates the $90\%$ credible interval of a fiducial population model from GWTC-4~\citep{GWTC-4_pop}.
  
In two-dimensional plots, the colored squares mark the positions of delta components.
Its color encodes the associated weight; lighter is higher weight and darker is lower weight.
We overlay the maximum-likelihood distribution of the parametric GWTC-4 population model in greyscale.  
The side panels contain the one-dimensional projections.

\begin{figure}[h]
  \centering  \includegraphics[width=0.45\textwidth]{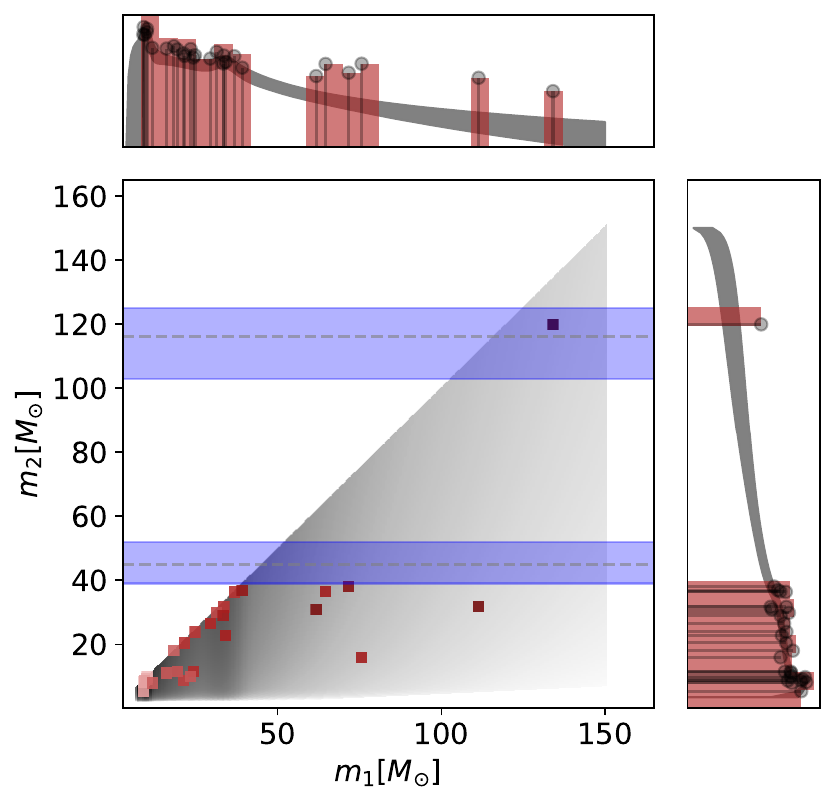}
  \caption{The $\pistroke$ distribution in the \((m_1,m_2)\) plane.  Red squares mark the delta-function components, with colour intensity proportional to their weights.  The grey shading represents the maximum-likelihood fit of the parametric GWTC-4 population model~\citep{GWTC-4}. A distinct secondary-mass gap appears for \(38\,M_\odot \lesssim m_2 \lesssim 120\,M_\odot\).  
  Horizontal blue bands show the gap boundaries inferred by~\citet{Tong_mass_gap}. The one-dimensional projection panels show the projected marginal distributions. }
  \label{fig:2d_m1_m2}
\end{figure}

\begin{figure}[h]
  \centering
  \includegraphics[width=0.45\textwidth]{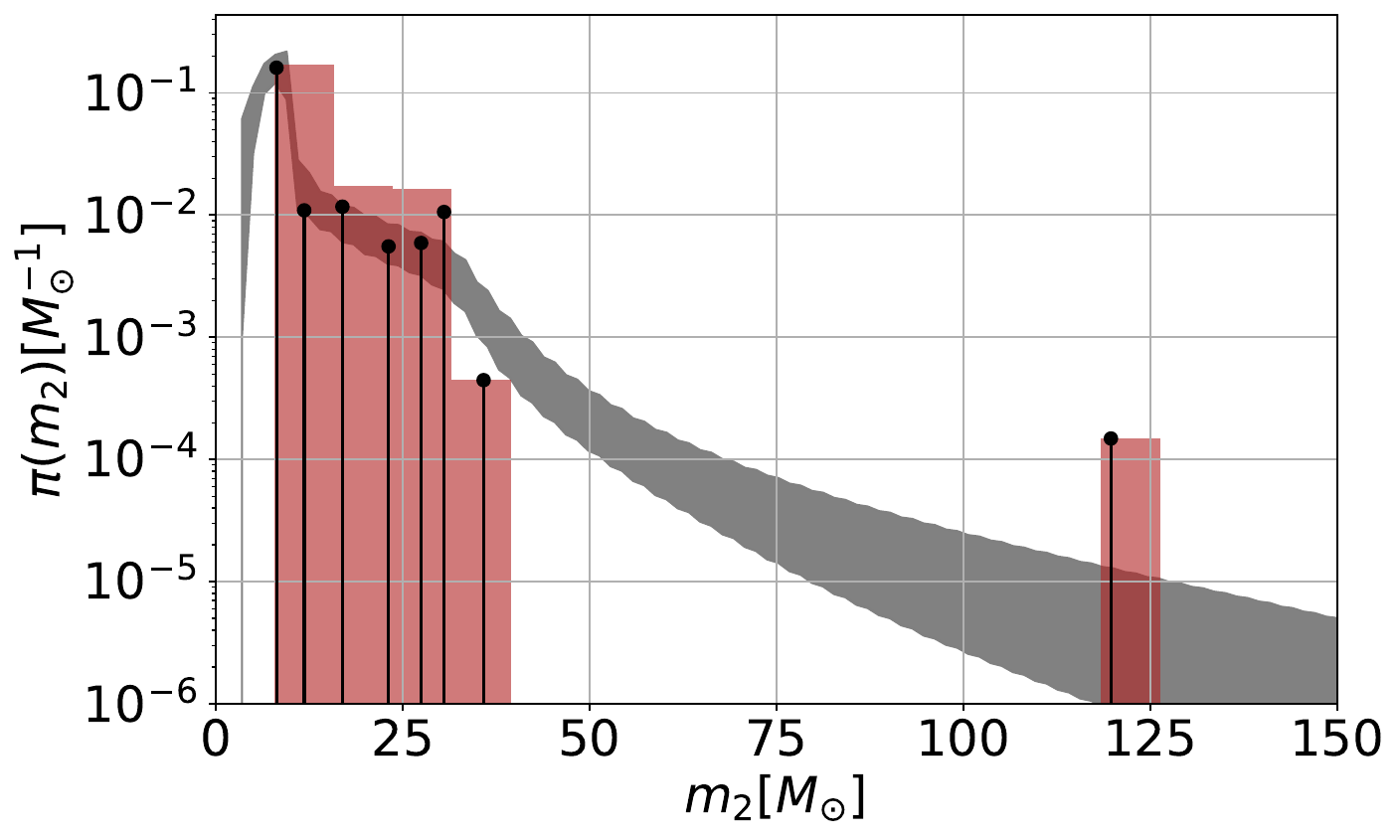}
  \caption{The $\pistroke$ distribution for \(m_2\). The black markers and vertical lines denote the delta-function components, with heights proportional to the corresponding weights. The red bars show a histogram of these components. A clear gap is evident between approximately \(38\,M_\odot\) and \(120\,M_\odot\).  The grey band indicates the $90\%$ credible interval of the parametric GWTC-4 population model from~\citet{GWTC-4}. }
  \label{fig:1d_m2}
\end{figure}

\begin{figure}[h]
  \centering
  \includegraphics[width=0.459\textwidth]{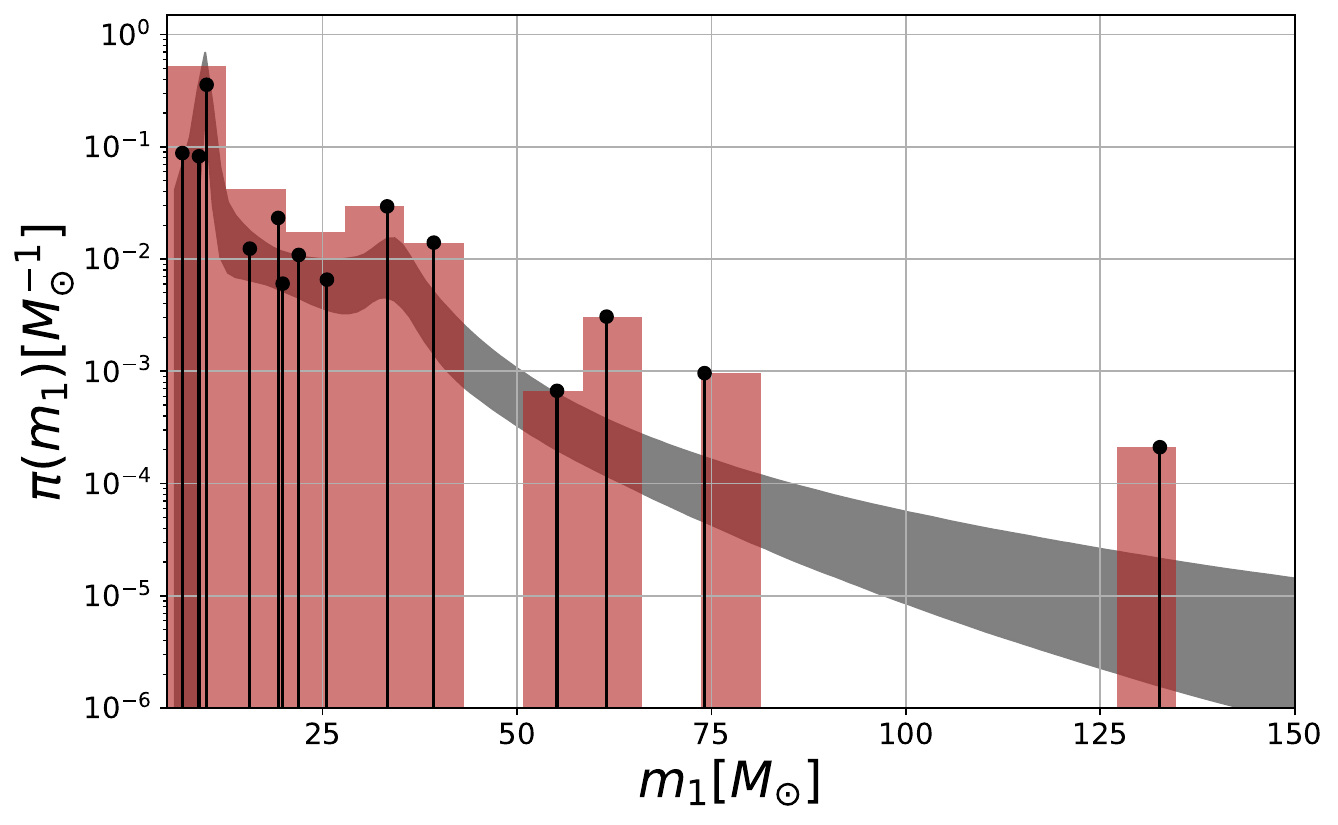}
  \caption{The $\pistroke$ distribution for \(m_1\). 
  The distribution is consistent with the GWTC-4 parametric population model and exhibits a excess near \(10\,M_\odot\), and \(35\,M_\odot\).}
  \label{fig:1d_m1}
\end{figure}

\subsection{Mass Properties}
\label{sec:results_mass}

Figure~\ref{fig:2d_m1_m2} reveals one of the most striking features of our analysis: a pronounced gap in the distribution of the secondary black-hole mass.
We identify a lower edge at \(m_2 \simeq 38\,M_\odot\) and an upper edge at \(m_2 \simeq 120\,M_\odot\); both edges are marked by $\pistroke$ components.
The feature is studied extensively in~\citet{Tong_mass_gap} using parametric models, which shows that the gap is statistically significant.
The gap is clearly visible in the one-dimensional \(m_2\) spectrum of Fig.~\ref{fig:1d_m2}, where the probability is strongly suppressed between \(38\,M_\odot\) and \(120\,M_\odot\)---in excellent agreement with the two-dimensional analysis.
The gap edges align closely with the values inferred by~\citet{Tong_mass_gap} with a parametric model; see the horizontal blue bands in Fig.~\ref{fig:2d_m1_m2}.
A plausible astrophysical interpretation for the gap is the onset of pair-instability supernovae, which inhibit the formation of black holes in this mass range~\citep[see, e.g.,][]{Woosley_2017,Farmer_2019}.  
We return to the implications of this hypothesis in Section~\ref{sec:discussion}.

The primary-mass distribution \(m_1\) (Fig.~\ref{fig:1d_m1}) broadly follows the population model and exhibits a clear excess around \(10\,M_\odot\) and \(35\,M_\odot\) in agreement with earlier studies~\citep{pi_stroke_orig,gwtc-3_pop,GWTC-4_pop}.  
Intriguingly, the \(m_2\) spectrum shows a less obvious excess near \(33\,M_\odot\) compared to the $m_1$ distribution. The proximity of these features suggests that both black-hole components associated with the excess may originate from the same formation pathway. This interpretation is reinforced by the preference for mass ratios $q \simeq 1$~\citep{midthirties}, visible as an accumulation along the $m_1 \simeq m_2$ diagonal near $35\,M_\odot$ in Fig.~\ref{fig:2d_m1_m2}.  The slightly lower value of the secondary mass can be explained by the labelling convention that designates the more massive object as $m_1$. 
A similar observation has been reported independently—using parametric modeling by~\citet{Farah_2024}, and a nonparametric kernel–density–estimation approach by~\citet{m2_mass_bump}.

kernel density estimation-based method

\begin{figure}[h]
  \centering
  \includegraphics[width=0.45\textwidth]{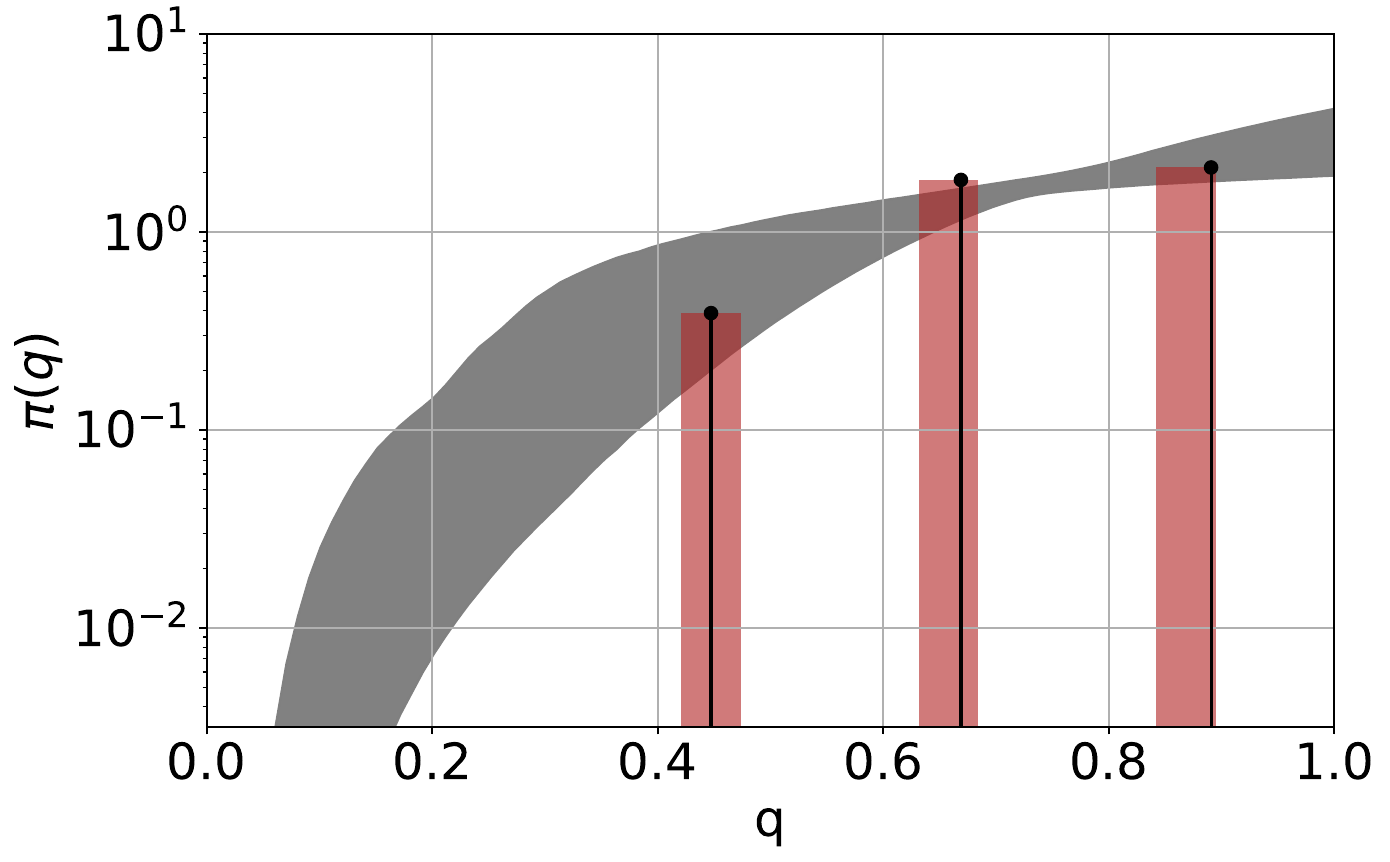}
  \caption{The $\pistroke$ distribution for mass ratio \(q=m_2/m_1\).}
  \label{fig:1d_q}
\end{figure}

Finally, the one-dimensional mass-ratio $\pistroke$ distribution shown in Fig.~\ref{fig:1d_q} follows slightly flatter distribution
compared to the parametric population model’s power-law~\citep{GWTC-4_pop}. This is consistent with the findings of~\citet{Fishbach_2020_mass_ratio} and~\citet{Golomb_mass_ratio}.

\subsection{Spin Properties}
\label{sec:results_spin}
\begin{figure}[h]
  \centering
  \includegraphics[width=0.45\textwidth]{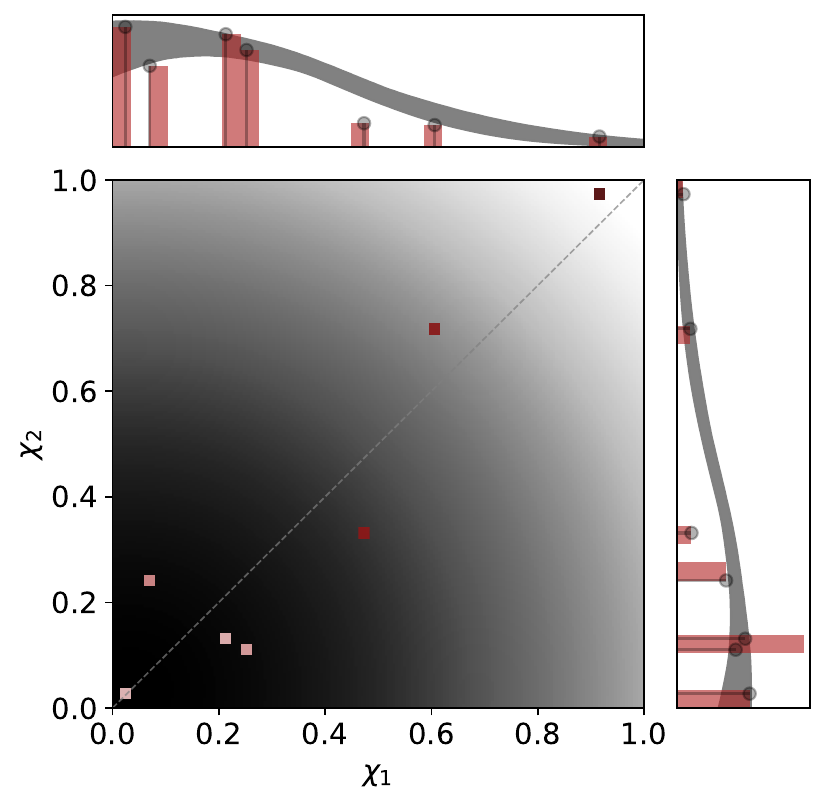}
  \caption{The $\pistroke$ distribution in the \((\chi_1,\chi_2)\) plane. The dashed diagonal line represents $\chi_1=\chi_2$.}
  \label{fig:2d_chi1_chi2}
\end{figure}

Figure~\ref{fig:2d_chi1_chi2}, shows the $\pistroke$ distribution in the (\(\chi_1,\chi_2\)) plane. 
We are struck by the apparent correlation
between $\chi_1$ and $\chi_2$.
While a large range of spin values are supported by the data, there is no support for large $\chi_1$ when $\chi_2$ is small and vice versa.
This result is consistent with~\citet{Adamcewicz_2025}, which suggests a $\chi_1, \chi_2$ correlation using a parameterized model.
This result is surprising since conventional thinking about binary evolution suggests that only one black hole is likely to have significant spin $\gtrsim 0.01$~\citep{MANDEL20221}
We discuss possible astrophysical implications in Sec.~\ref{sec:discussion}.   

The joint distribution is suggestive of a peak near \(\chi_1 \simeq \chi_2 \simeq 0.2\) and appears to taper smoothly toward higher spins.  
The corresponding one-dimensional distributions shown in Figs.~\ref{fig:1d_chi1} and~\ref{fig:1d_chi_p_chi2} (see Appendix~\ref{appx:additional__pi_stroke}) broadly follow the population model and display an excess at \(\chi_1 \approx 0.16\) and \(\chi_2 \approx 0.13\).
These results are also consistent with~\citet{Adamcewicz_2025}, which uses a parameterised model to find support for sub-populations with $\chi\approx0.2$ and $\chi\approx0.8$.

We note that the $\chi_2$ marginal obtained from the two-dimensional ($\chi_1$, $\chi_2$) analysis differs modestly from the standalone one-dimensional $\chi_2$ distribution. This hints that the joint distribution is not separable, i.e., $\chi_1$ and $\chi_2$ are statistically dependent. Further details are provided in Appendix~\ref{appx:compare_1D_to_2D}.

\begin{figure}[h]
  \centering
  \includegraphics[width=0.45\textwidth]{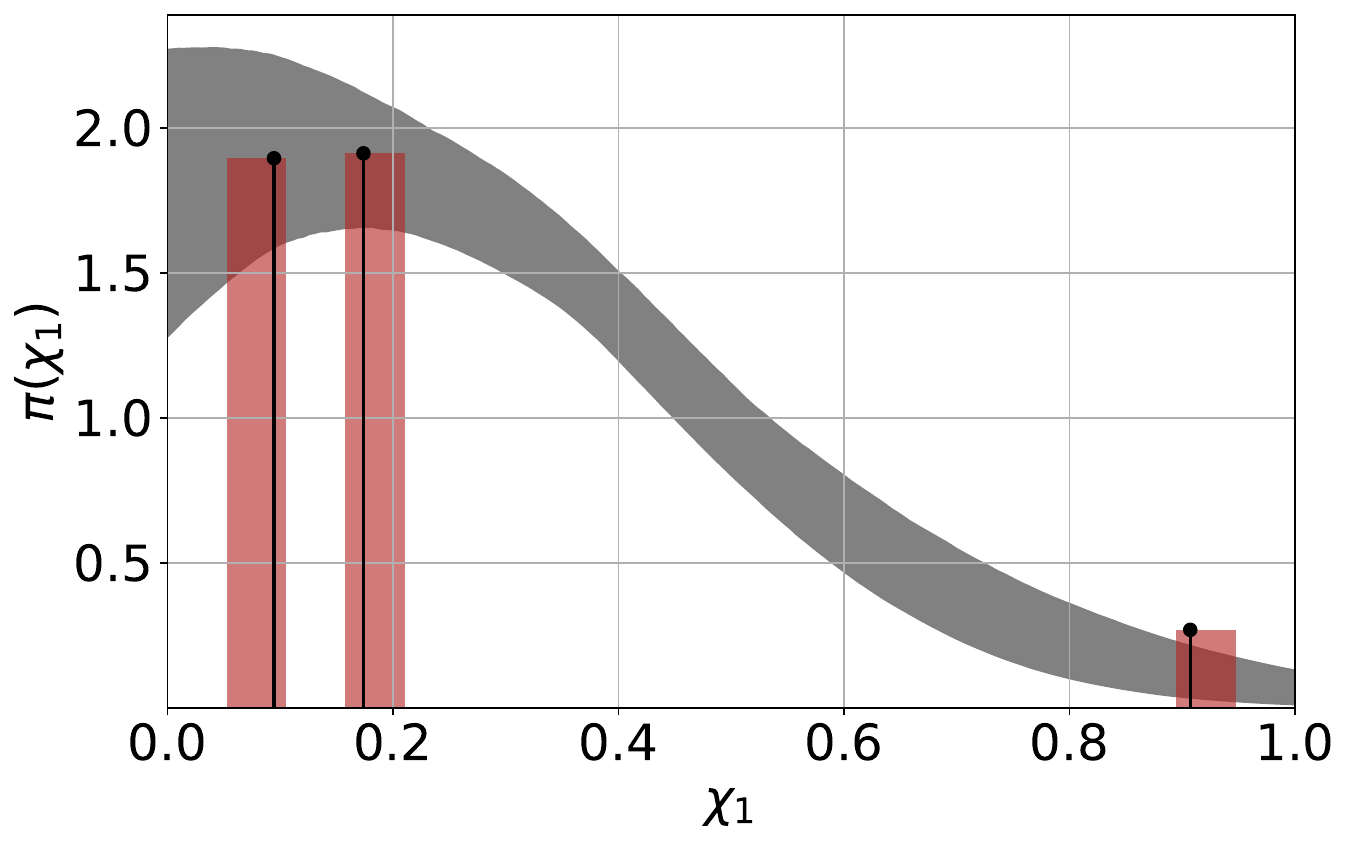}
  \caption{The $\pistroke$ distribution for spin magnitude \(\chi_1\).}
  \label{fig:1d_chi1}
\end{figure}

\begin{figure}[h]
  \centering
  \includegraphics[width=0.45\textwidth]{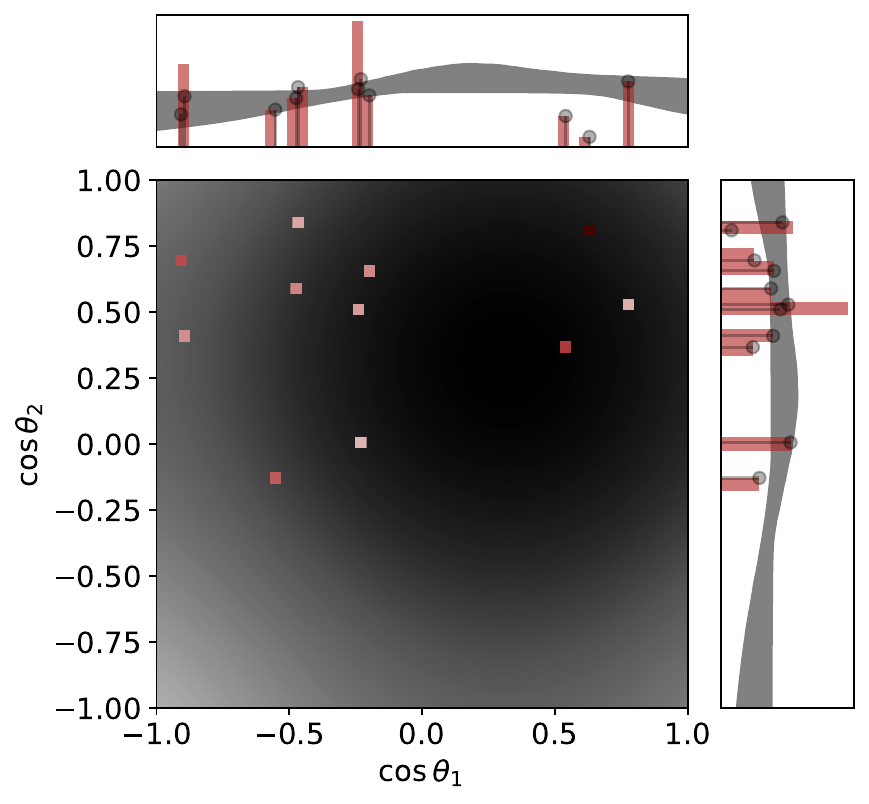}
  \caption{The $\pistroke$ distribution in the \((\cos\theta_1,\cos\theta_2)\) plane. }
  \label{fig:2d_cos1_cos2}
\end{figure}

\begin{figure}[h]
  \centering
  \includegraphics[width=0.45\textwidth]{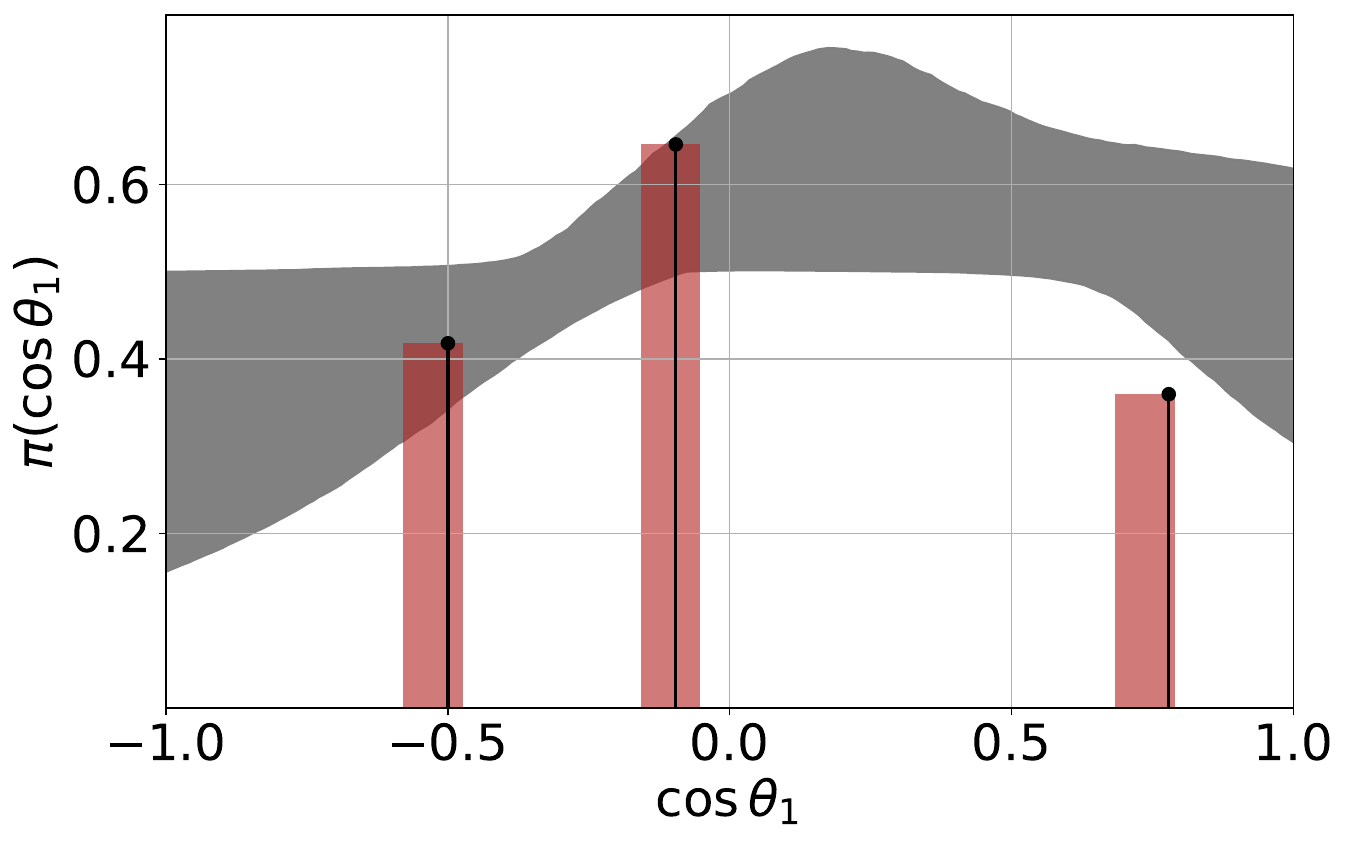}
  \caption{The $\pistroke$ distribution for  \(\cos\theta_1\). }
  \label{fig:1d_cos1}
\end{figure}

\begin{figure}[h]
  \centering
  \includegraphics[width=0.45\textwidth]{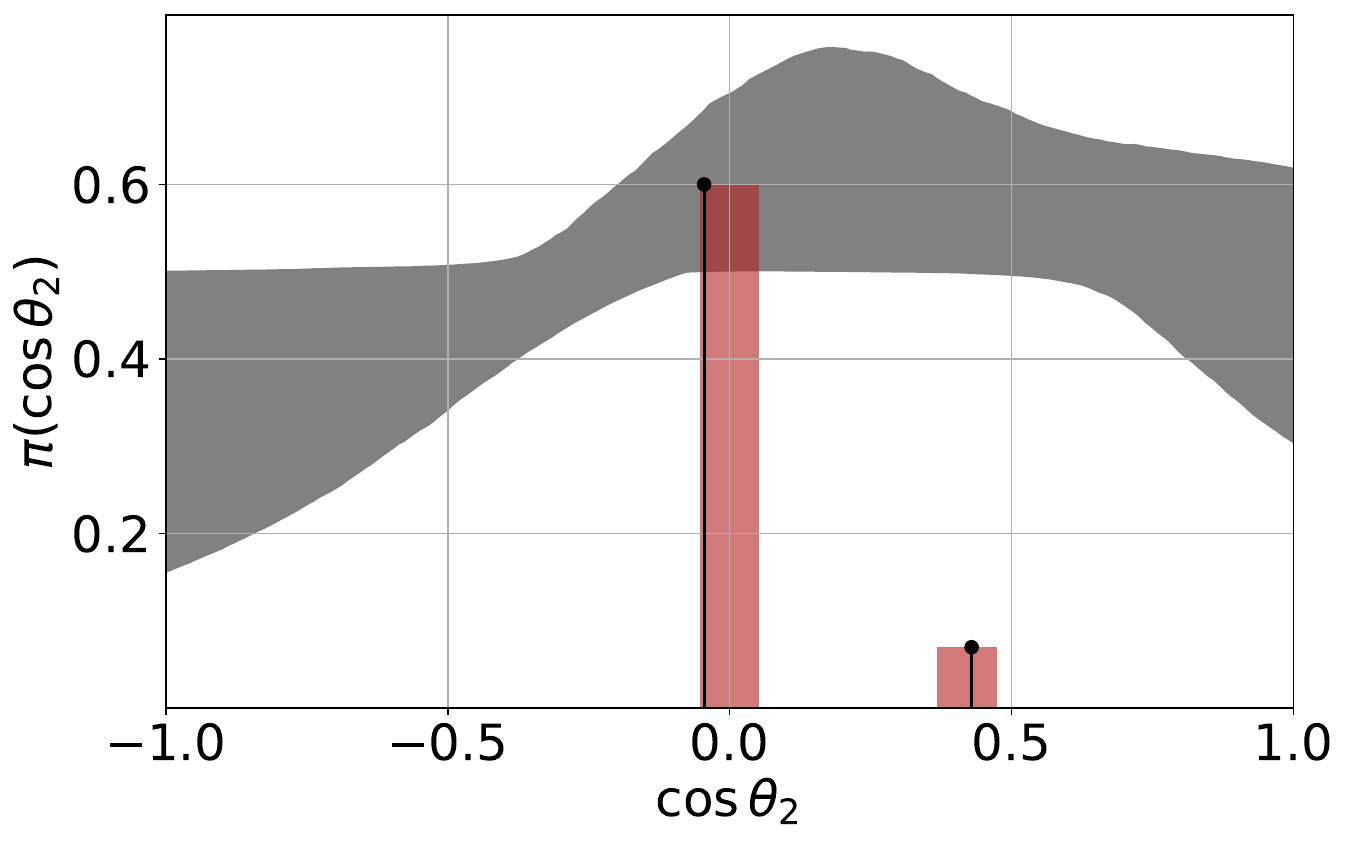}
  \caption{The $\pistroke$ distribution for \(\cos\theta_2\).}
  \label{fig:1d_cos2}
\end{figure}

The two-dimensional cosine tilt–angle distribution in Fig.~\ref{fig:2d_cos1_cos2} reveals an intriguing  pattern. Whereas \(\cos\theta_1\) is nearly uniform, \(\cos\theta_2\) shows a clear preference for aligned spins (\(\cos\theta_2 > 0\)).  This two-dimensional behavior is consistent with the one-dimensional $\pistroke$ distributions shown in Figs.~\ref{fig:1d_cos1} and~\ref{fig:1d_cos2}, yet it differs from the GWTC-4 parametric population model, which assumes identical tilt-angle distributions for the two black holes. 
Using only the $\pistroke$ formalism alone, it is impossible to assess if this trend is statistically significant, or merely a result of noise fluctuations.
However, in Sec.~\ref{sec:discussion} we speculate on potential astrophysical origins of this feature.

\begin{figure}[h]
  \centering
  \includegraphics[width=0.45\textwidth]{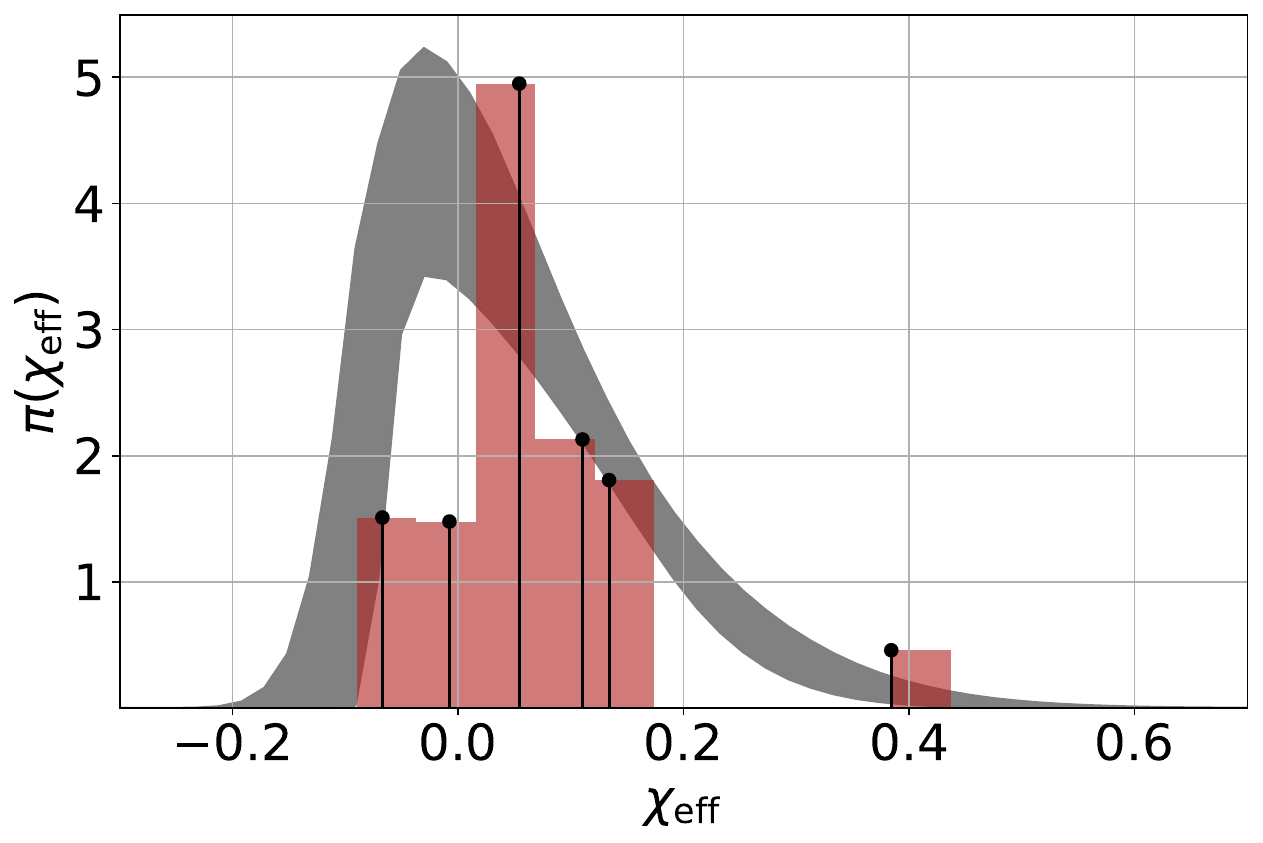}
  \caption{The $\pistroke$ distribution for $\chi_{\mathrm{eff}}$.}
  \label{fig:1d_chi_eff}
\end{figure}

Figure~\ref{fig:1d_chi_eff} displays the one-dimensional $\pistroke$ reconstruction of the effective inspiral spin parameter \(\chi_{\mathrm{eff}}\).  The shape broadly tracks the GWTC-4 parametric model, showing modest support for negative values but a pronounced skew toward positive \(\chi_{\mathrm{eff}}\) values, in agreement with the findings of~\citet{banagiri_2025}. Compared with the parametric fit, the $\pistroke$ result assigns less probability to the negative tail.   
This finding is broadly consistent with the earlier $\pistroke$ analysis of \(\chi_{\mathrm{eff}}\), which found virtually no support for negative values~\citep{pi_stroke_orig}.

Figure~\ref{fig:1d_chi_p_chi2} (see Appendix \ref{appx:additional__pi_stroke}) presents the $\pistroke$ estimate for \(\chi_{\mathrm{p}}\).  All probability mass collapses into a single delta function underscoring that current data provides very little constraining power for this parameter. 
To test robustness, we removed two events reported to have high $\chi_{\mathrm{p}}$ values (\textsc{GW200129}~\citep{GWTC-3} and \textsc{GW231123}~\citep{gw231123}) and re-ran the $\pistroke$ analysis; we again recovered a single delta function at the same location. Although the resolving power is weak, this result is indicates mild population-level support for nonzero $\chi_{\mathrm{p}}$.

\subsubsection{Spin-Mass Properties}
\label{subsec:results_spin_mass}

\begin{figure}[h]
  \centering
  \includegraphics[width=0.45\textwidth]{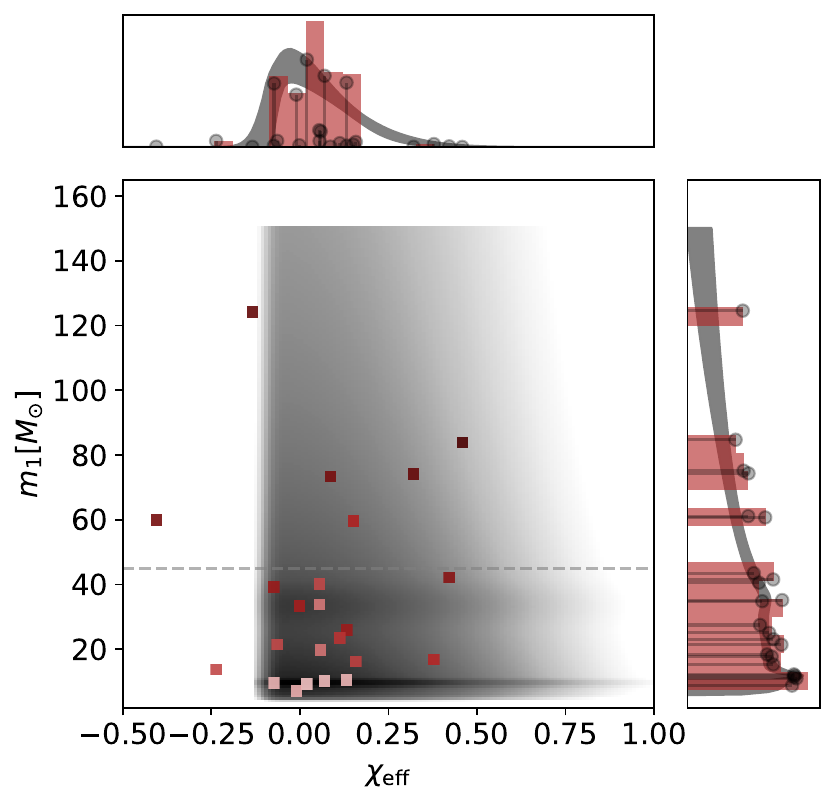}
  \caption{The $\pistroke$ distribution in the \((\chi_{\mathrm{eff}},m_1)\) plane.
The dashed horizontal line at \(m_1 = 45\,M_\odot\) marks the transition
identified by~\citet{antonini_chi_eff}; see also~\citet{Tong_mass_gap}. }
  \label{fig:2d_m1_chi_eff}
\end{figure}

Figure~\ref{fig:2d_m1_chi_eff} displays the
$\pistroke$ reconstruction in the \((\chi_{\mathrm{eff}},m_1)\) plane. The delta–function components reveal a trend first noted by~\citet{antonini_chi_eff}: for binaries with primary black hole mass
\(m_1 < 45\,M_\odot\), the effective–spin distribution appears to be narrowly
peaked around \(\chi_{\mathrm{eff}}\approx0\) while above this threshold the
distribution appears to broadens.  
In addition, our two–dimensional
analysis recovers appreciable support for negative
\(\chi_{\mathrm{eff}}\) values, in contrast to the one–dimensional
result of Fig.~\ref{fig:1d_chi_eff}.

\begin{figure}[h]
  \centering
  \includegraphics[width=0.45\textwidth]{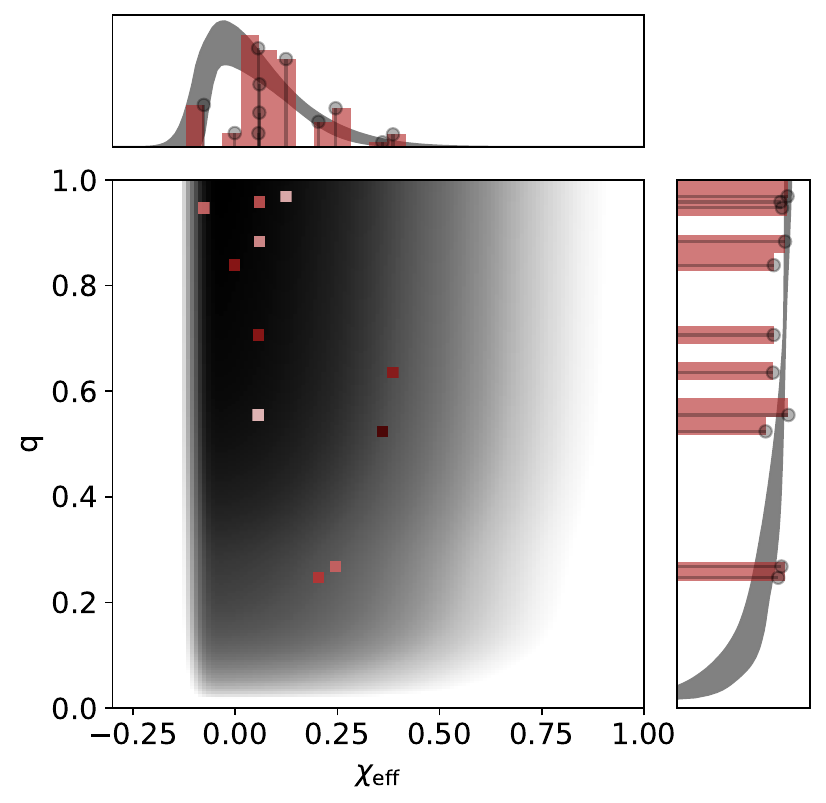}
  \caption{The $\pistroke$ distribution in the \((\chi_{\mathrm{eff}}, q)\) plane.}
  \label{fig:2d_q_chi_eff}
\end{figure}

Figure~\ref{fig:2d_q_chi_eff} shows the corresponding
\((\chi_{\mathrm{eff}}, q)\) reconstruction. As reported in earlier studies~\citep{Callister_2021,Christian_q_chi_eff}, we recover an apparent anti-correlation: 
systems with smaller mass ratios show an increasing trend toward larger positive 
effective spins.  Although this trend matches previous
findings, we caution that it may arise from the unmodeled dependence
of \(\chi_{\mathrm{eff}}\) on \(m_1\). We discuss this possibility in
Sec.~\ref{sec:discussion}.     

\subsection{Redshift Properties}
\label{sec:results_redshift}
\begin{figure}[h]
  \centering
  \includegraphics[width=0.45\textwidth]{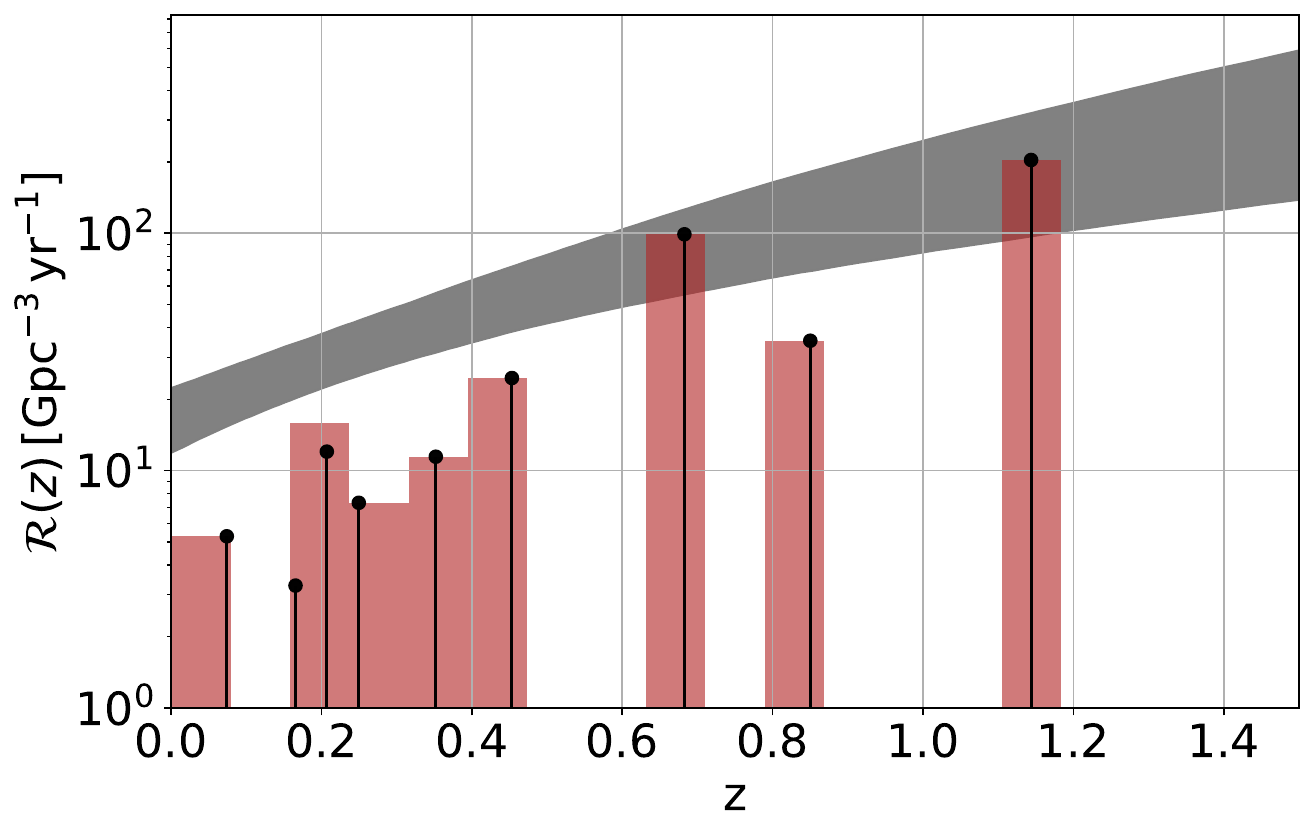}
  \caption{The $\pistroke$ reconstruction of the merger rate as a function of redshift, \(z\). }
  \label{fig:1d_z}
\end{figure}

Figure \ref{fig:1d_z} shows the one-dimensional $\pistroke$ reconstruction of the merger-rate evolution with redshift; the conversion from a distribution to a rate is described in Appendix~\ref{appx_subsec:redshift_evolution}.
The data-driven estimate broadly tracks the parametric GWTC-4 population model, but it departs from the model at low redshifts, a discrepancy also noted in the previous $\pistroke$ study~\citep{pi_stroke_orig}.  

\subsubsection{Redshift-Spin Properties}
\label{subsec:results_redshift_spin}
\begin{figure}[h]
  \centering
  \includegraphics[width=0.45\textwidth]{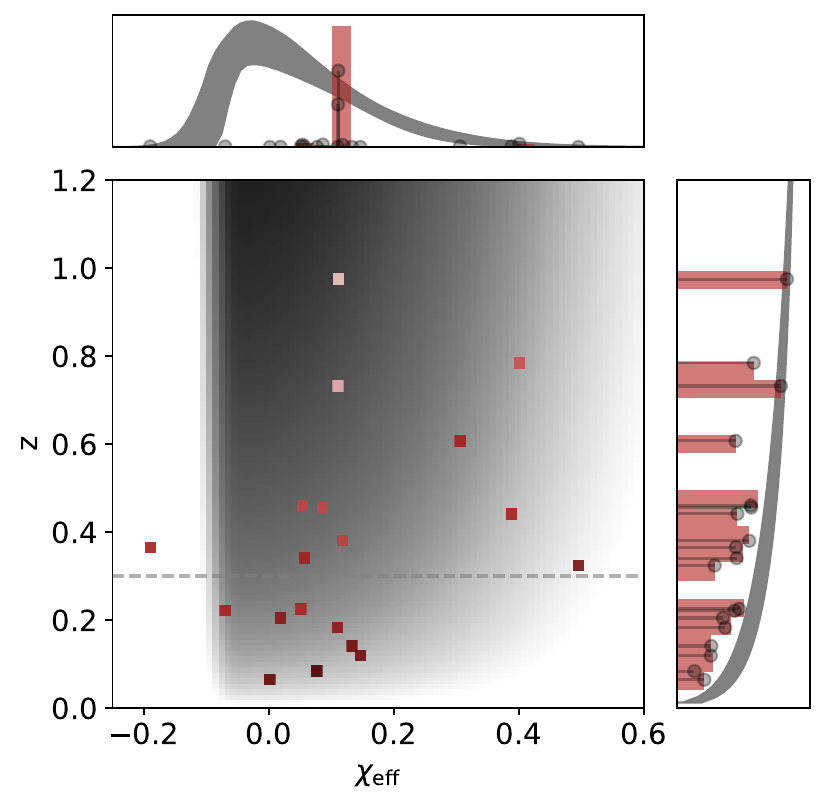}
  \caption{The $\pistroke$ distribution in the \((\chi_{\mathrm{eff}},z)\) plane. The dashed horizontal line at \(z=0.3\) marks the estimated spin-transition redshift. Note that the $z$ probability distribution shown here incorporates both the differential comoving volume and the merger-time delay, unlike Fig.~\ref{fig:1d_z}. }
  \label{fig:2d_chi_eff_z}
\end{figure}

Figure~\ref{fig:2d_chi_eff_z} presents the two–dimensional
$\pistroke$ reconstruction in the \((\chi_{\mathrm{eff}},z)\) plane. Similarly to the \((m_1,\chi_{\mathrm{eff}})\) analysis, the $\chi_{\mathrm{eff}}$ distribution width increases with redshift, where the transition seem to occur around $z\sim0.3$. This is most likely the result of selection effects that induce an $m_1$–$z$ correlation. Related correlations involving $m_1, \, z$, and $\chi_{\mathrm{eff}}$ were reported by \citet{Biscoveanu_2022}. 
Surprisingly, the marginal $\chi_{\mathrm{eff}}$ is characterised by a sharp feature at $\chi_\text{eff}\approx0.15$, which we do not see in the one-dimensional $\pistroke(\chi_\text{eff})$.
This occurs because selection effects require a few high-\(z\) components (see Fig.~\ref{fig:1d_z}), those components are free to concentrate at the significant $\chi_{\text{eff}}$ mode (see Fig.~\ref{fig:1d_chi_eff}). We elaborate on this in Sec.~\ref{sec:discussion}.   

\subsection{Informativeness}

\begin{table}[t]
  \centering
  \caption{Informativeness (\(\mathcal{I}=N_{\delta /153}\)) recovered in each $\pistroke$ analysis.}
  \label{tab:informative_fraction}
  \begin{tabular}{l@{\hspace{1.2em}}r |
                  l@{\hspace{2.2em}}r}
    \toprule
    $\pistroke$ Analysis & \(\mathcal{I}\)  & $\pistroke$ Analysis & \(\mathcal{I}\) \\
    \hline
    $m_1$                       & 14/153   & $\cos\theta_1$                &  3/153 \\
    $m_2$                         &  8/153   & $\cos\theta_2$      &  2/153 \\
    $(m_1,m_2)$                  &  27/153   & $\chi_{\mathrm{eff}}$          &  6/153 \\
    $q$                    &  3/153   & $\chi_{\mathrm{p}}$           &  1/153 \\
    $(\chi_1,\chi_2)$     &  7/153   & $(\chi_{\mathrm{eff}},m_1)$          &  23/153 \\
    $\chi_1$             &  3/153   & $(\chi_{\mathrm{eff}}, q)$   &  11/153 \\
    $\chi_2$             & 2/153        & $z$    & 9/153      \\
    $(\cos\theta_1,\cos\theta_2)$     &  11/153   & $(\chi_{\mathrm{eff}},z)$               &  19/153 \\
    \hline
  \end{tabular}
\end{table}

Table~\ref{tab:informative_fraction} lists the informativeness
\(\mathcal{I}\) recovered in each $\pistroke$ analysis (see Eq.~\eqref{eq:informativeness}). Two factors determine the value of \(\mathcal{I}\):
\begin{enumerate}
  \item \textbf{Intrinsic structure.}  Distributions with sharp edges or multiple modes require more delta functions to reproduce, and therefore yield larger \(\mathcal{I}\).

  \item \textbf{Measurement precision.}  Tighter posteriors act like higher‐resolution pixels, resolving finer-features of the underlying distribution and thus demanding a denser set of delta components, increasing \(\mathcal{I}\).
\end{enumerate}
Because a two–dimensional analysis can expose richer structure than a
one–dimensional projection, the two-dimensional analyses tend to produce larger values of ${\cal I}$, all else being equal.

\section{Discussion}
\label{sec:discussion}
\subsection{Black hole masses}
The $\pistroke$ distributions for GWTC-4 yield a number of astrophysical insights.
In Figs~\ref{fig:2d_m1_m2}-\ref{fig:1d_m2}, we observe a gap in the secondary mass spectrum consistent with the gap identified by~\citet{Tong_mass_gap} with a parameterised model.
No delta functions are assigned for masses in the range $m_2 \simeq 40\text{--}120\,M_\odot$.
The gap is consistent with theoretical predictions for the pair-instability supernova ~\citep[PISN; ][]{Woosley_2017,Farmer_2019}. Black holes formed from isolated stellar evolution are expected to be absent in this mass interval, as pair-instability processes either completely disrupt the progenitor star or cause substantial mass loss, resulting in remnants below approximately $\sim 45\,M_\odot$.

However, the question remains: why do we observe a gap in $m_2$, but not $m_1$?\footnote{We note two apparent gaps in $m_1$: $\sim40\text{--}60\,M_\odot$ and $\sim75\text{--}110\,M_\odot$, but without a supporting parametric model, we cannot assess their statistical significance. We also note that one expects a gap in $m_1$ between $2M_{\rm low}$ and $M_{\rm up}$ if first–generation black holes are bounded below the lower edge $M_{\rm low}$ of the PISN gap and above the upper edge is $M_{\rm up}$; numerically, this corresponds to $\sim80\text{--}120\,M_\odot$.}
One possible explanation is that binary black holes with primary mass above $45 M_\odot$ are remnants of previous mergers---second-generation (2G) black holes, paired predominantly with first-generation (1G) companions. 
Since 2G black holes are not formed through stellar evolution, they can populate the pair instability gap while their 1G companions cannot.
Contribution from 2G+2G mergers are predicted to be relatively rare~\citep{PhysRevD.100.041301,PhysRevD.100.043027,refId0}, though, one would expect some level of contamination to eventually emerge in the $m_2$ mass gap from 2G+2G mergers. 

Further support for this scenario is provided by our joint $( \chi_{\mathrm{eff}},m_1)$ analysis, which suggests a broadening of the $\chi_{\mathrm{eff}}$ distribution above $\sim45\,M_\odot$ (Fig.~\ref{fig:2d_m1_chi_eff}), consistent with recent findings by~\citet{antonini_chi_eff}, and seen with greater statistical significance in~\citet{Tong_mass_gap}.
We interpret the black holes below $45 M_\odot$ as predominantly 1G, which are expected to have comparatively small dimensionless spins~\citep{Fuller_Ma}.
On the other hand, black holes above $45 M_\odot$ we interpret as 2G, which inherit the angular momentum of their parent binary leading to larger spins and a wider $\chi_\text{eff}$ distribution~\citep{GerosaFishbach}.

Another consequence of the broadening we observe in the $(\chi_{\mathrm{eff}},m_1)$ plane is that the apparent anti-correlation between mass ratio $q$ and $\chi_{\mathrm{eff}}$ (Fig.~\ref{fig:2d_q_chi_eff}) might arise as a projection effect.
At primary masses \(m_1 \gtrsim 45\,M_\odot\) the $\chi_{\mathrm{eff}}$ distribution broadens. Since the pair-instability gap does not allow for secondaries with \(m_2 \gtrsim 45\,M_\odot\),
binaries in this high-mass regime must have lower mass ratios.
Thus, when one analyses $\chi_{\mathrm{eff}}$ versus $q$, the broader spin distribution at low $q$ mimics anti-correlation, even if $\chi_{\mathrm{eff}}$ and $q$ are actually independent once $m_1$ is taken into account.

This interpretation is supported by our one-dimensional $\chi_{\mathrm{eff}}$ reconstruction (Fig.~\ref{fig:1d_chi_eff}), which shows little support for negative values---calculated assuming 
\textsc{Power Law + Peak}~\citep{Talbot_2018} (See. Appendix~\ref{appx:pop_model})---while negative $\chi_{\mathrm{eff}}$ arises when $m_1$ is included as an explicit $\pistroke$ parameter.
We hypothesise that a three-dimensional analysis of ($q,\chi_{\mathrm{eff}},m_1$) would recover the same broadening at low $q$ while eliminating the apparent anti-correlation. This interpretation also aligns with the absence of a correlation reported by~\citet{midthirties}, who allowed for dependencies between $m_1$, and $m_2$ in their ($\chi_{\mathrm{eff}}, q$) analysis. \citet{chi_eff_mass_ratio_anti_anti_correlation} reconstruct the joint $(m_1,m_2,\chi_{\mathrm{eff}})$ distribution and motivate a three-dimensional analysis to disentangle population structure.
We defer such an investigation to future work.

The excess of events around a primary mass of $\sim 35\,M_\odot$ persists (Fig.~\ref{fig:1d_m1}), with possible hints of a similar excess in the secondary mass distribution (Fig.~\ref{fig:1d_m2}). However, given that the pair-instability mass gap is expected to begin near $\sim 45\,M_\odot$, it seems unlikely that this excess is caused by pair-instability supernovae, as argued by~\citet{bump_mass_simulation,bump_mass_simulation2,bump_mass,midthirties}.

\subsection{Black hole spin}
Our $\pistroke$ hints at correlated structure in the \((\chi_1,\chi_2)\) plane (Fig.~\ref{fig:2d_chi1_chi2}).
The largest weights cluster at low values \(\chi_1 \simeq \chi_2 \simeq 0.1\).  
This two-dimensional pattern agrees with the one-dimensional peak we recover at \(\chi \sim 0.2\). 
This apparent correlation between \(\chi_1\) and \(\chi_2\) hints at a lack of binaries in which only one component carries substantial spin.  
However, it could also be a manifestation of Simpson’s paradox whereby several latent groups, each with distinct properties, combine to mimic an intrinsic \(\chi_1\)–\(\chi_2\) correlation in the aggregate data~\citep{simpson}.
These same patterns are found in a parametric study of GWTC-4 by~\citet{Adamcewicz_2025}.

These results are all consistent with recent population analyses using GWTC-3 data.
Various highly-parameterised reconstructions report a broad peak at \(\chi\!\approx\!0.2\)~\citep{Golomb_Talbot_data_driven,Callister_data_driven} and find no evidence for a spike at \(\chi = 0\)~\citep{Tong_non_zero}. 
Other analyses require a dominant low-spin group together with a smaller fraction of systems in which both black holes are spun-up~\citep{hussain_magnitude_correlations}, while models that allow just a single rapidly rotating component are disfavored~\citep{Adamcewicz_2024}.  

As the specifics of angular-momentum transport in high-mass stars remain uncertain, it is commonly supposed that stellar cores lose most of their angular momentum prior to collapse, producing black holes with negligible spins~\citep{Fuller_Ma}.  On the other hand, mergers of non-spinning binary black holes with approximately equal mass are expected to produce a second-generation black hole with \(\chi \sim 0.7\)~\citep{Fishbach_2017}. 
One would therefore expect 1G+2G binaries to merge with $\chi_1\approx 0.7$ and $\chi_2\lesssim0.2$.
The lack of $\pistroke$ support for such systems is therefore surprising.
We speculate that delta functions with $\chi_1\approx 0.7$ and $\chi_2\lesssim0.2$ might begin to appear with the addition of more data perhaps because it is difficult to measure $\chi_2$ in 1G+2G mergers when the mass ratio is typically $q\lesssim0.5$.
We hope to investigate this hypothesis with future work.

The preference for $\cos \theta_2 > 0$ (Figs.~~\ref{fig:2d_cos1_cos2} and \ref{fig:1d_cos2}) hint that secondary black hole spins may be preferentially aligned with the orbital angular momentum.
However, the statistical significance of this feature is not yet assessed with a parameterized model.
If this trend is real, it may be an imprint of binary evolution. One plausible isolated binary evolution scenario outlined by~\citet{Gerosa2013} involves significant natal kicks tilting the orbit at the first supernova. 
However, the secondary is still a star, and efficient tidal interactions during mass transfer or common-envelope evolution can realign its spin with the new orbit, while the primary black hole remains misaligned.
When the second black hole forms, its natal kick further perturbs the orbit, typically more mildly, leading to a typical configuration where $\theta_1 > \theta_2$. If the mass ratio remains unreversed, this asymmetry can persist, yielding the observed $\cos \theta_1 < \cos \theta_2$. This picture is expanded further  in~\citet{Stevenson2017} and~\citet{Gerosa2018}.

We caution that this feature may be influenced by simplifying assumptions, e.g., treating $\chi_1$ and $\chi_2$ as independent, despite the data indicating a correlation (see Fig.~\ref{fig:2d_chi1_chi2}). Extending parametric models to allow correlated spin magnitudes and component-dependent tilt distributions would provide a direct test of this trend in future work.

\subsection{Black hole redshift}
The redshift distribution is qualitatively similar to the parametric population model in both the one-dimensional $\pistroke$ analysis (Fig.~\ref{fig:1d_z}) and the two-dimensional ($\chi_{\mathrm{eff}}$, $z$) distribution (Fig.~\ref{fig:2d_chi_eff_z}).  The apparent broadening of the $\chi_{\mathrm{eff}}$ distribution at higher redshift could arise from an unmodelled correlation with the primary mass $m_1$.
More distant detections tend to involve higher-mass systems, because higher-mass binaries are louder and thus more easily detected at larger distances, which based on the ($\chi_{\mathrm{eff}},m_1$) $\pistroke$ analysis is expected to broaden the range of $\chi_{\mathrm{eff}}$ values. 

The narrow feature at $\chi_{\text{eff}}\!\approx\!0.15$ seen in the $(\chi_{\text{eff}},z)$ $\pistroke$ analysis likely reflects a projection effect driven by selection effects. Only a handful of events have a maximum-likelihood redshifts at high values (\(z\gtrsim 0.8\)), yet the intrinsic merger rate increases with redshift, implying that the detection probability is small there. Consequently, to account for selection effects, the $\pistroke$ fit assigns substantial weight to a few high-$z$ delta-function components. Because there are only a few gravitational-wave events in that region that are not well resolved, those high-weight components are optimized toward a significant mode of the $\chi_{\text{eff}}$ distribution (see Fig.~\ref{fig:1d_chi_eff}), producing an artificial spike in the $\chi_{\text{eff}}$ marginal distribution. 
To support this interpretation, we present two pieces of evidence: (i) prior work indicates that $\chi_{\text{eff}}$ is correlated either with $z$ or with $m_1$ (or both), indicating an observational coupling among \((m_1,\, z,\, \chi_{\text{eff}})\)  ~\citep{Biscoveanu_2022}; consistently, our $(\chi_{\text{eff}},m_1)$ analysis (see Fig.~\ref{fig:2d_m1_chi_eff}) shows that the $\chi_{\text{eff}}$ distribution changes above $m_1\!\approx\!45\,M_\odot$ relative to the low-mass regime, so one expects a corresponding change at higher redshift, i.e., the $(\chi_{\text{eff}},z)$ marginal need not match the one-dimensional $\pistroke(\chi_{\text{eff}})$  distribution, and (ii) removing several of the largest high-$z$ delta functions (or the top few by weight) causes the $\chi_{\text{eff}}$ marginal in the $(\chi_{\text{eff}},\,z)$ analysis to converge toward the one-dimensional $\pistroke(\chi_{\text{eff}})$ result.

\section{Conclusions and public data}
\label{sec:conclusions}
We apply  the $\pistroke$ methodology to GWTC-4.
The $\pistroke$ formalism proves a versatile tool.
First, it verifies and validates established population trends, as illustrated by the primary and secondary mass spectra, both of which exhibit an excess near $\sim 35 M_{\odot}$. The one-dimensional spin distribution follows the GWTC-4 parametric model, as does the redshift distribution.
Second, it flags possible model misspecifications: for example, the secondary-mass spectrum clearly reveals a mass gap, possibly associated with pair-instability supernovae and recently identified by \citet{Tong_mass_gap}.
Third, it spotlights features preferred by the data, such as a possible \(\chi_1\! \simeq\!\chi_2\) correlation \citep{Adamcewicz_2025}, a possible preference for $\cos\theta_2 >0$, and perhaps a three-dimensional correlation between $m_1,\,q, \chi_\text{eff}$.
These emerging features can help guide the development of next-generation population models.

Finally, $\pistroke$ samples are useful for theorists who want to compare predictions to LVK data without assuming a specific parameterised fit.
Plotting predictions alongside $\pistroke$ samples is generally more useful than plotting predictions alongside the maximum-likelihood point estimate for each event. 
In the limit that the number of measurements $N$ becomes large, the $\pistroke$ samples converge to the true astrophysical distribution whereas the distribution of maximum-likelihood samples does not.
Also, the fact that $\pistroke$ is characterized by only $n$ delta functions (with $n<N$), provides a more indicative visualization of the information in the gravitational-wave catalog than a plot of maximum-likelihood values, which can provide a misleading picture of what we know about the distribution of different parameters.
Thus, all of the $\pistroke$ samples for this paper are available on Zenodo at the following link~\citet{pi_stroke_zenodo}.

\section*{Acknowledgements}

This work is supported through the Australian Research Council (ARC) Centre of Excellence CE230100016, Discovery Projects DP220101610 and DP230103088, and LIEF Project LE210100002.
We are grateful to Simon Stevenson, Christian Adamcewicz, and Sharan Banagiri for insightful discussions. We thank Jack Heinzel and Thomas Dent for feedback on an earlier draft of this manuscript.

This material is based upon work supported by NSF's LIGO Laboratory which is a major facility fully funded by the National Science Foundation.
The authors are grateful for computational resources provided by the LIGO Laboratory and supported by National Science Foundation Grants PHY-0757058 and PHY-0823459.

This research has made use of data or software obtained from the Gravitational Wave Open Science Center (gw-openscience.org), a service of LIGO Laboratory, the LIGO Scientific Collaboration, the Virgo Collaboration, and KAGRA. LIGO Laboratory and Advanced LIGO are funded by the United States National Science Foundation (NSF) as well as the Science and Technology Facilities Council (STFC) of the United Kingdom, the Max-Planck-Society (MPS), and the State of Niedersachsen/Germany for support of the construction of Advanced LIGO and construction and operation of the GEO600 detector. Additional support for Advanced LIGO was provided by the Australian Research Council. Virgo is funded, through the European Gravitational Observatory (EGO), by the French Centre National de Recherche Scientifique (CNRS), the Italian Istituto Nazionale di Fisica Nucleare (INFN) and the Dutch Nikhef, with contributions by institutions from Belgium, Germany, Greece, Hungary, Ireland, Japan, Monaco, Poland, Portugal, Spain. The construction and operation of KAGRA are funded by Ministry of Education, Culture, Sports, Science and Technology (MEXT), and Japan Society for the Promotion of Science (JSPS), National Research Foundation (NRF) and Ministry of Science and ICT (MSIT) in Korea, Academia Sinica (AS) and the Ministry of Science and Technology (MoST) in Taiwan.

\appendix

\section{Normalizing Flows and Kernel Density Estimator}
\label{appx:nf_method}

Normalising flows provide a powerful way to approximate an
arbitrary probability density from a finite set of samples~\citep{JMLR:v22:19-1028}.  
Here we use normalising flows both for the marginal–likelihood evaluation
(Section~\ref{sec:marg_like_subset}) and for modelling the empirical
detected injections distribution (Section~\ref{sec:selection_effects}).  
All flows are implemented with the \textsc{FlowJax}
library~\citep{flowjax-doc}.

For one-dimensional marginal-likelihood evaluations, we use the kernel density estimator implemented in SciPy~\citep{scott1992multivariate,Silverman_kde}.

The main ingredients are summarised below.

\subsubsection*{1. Pre-processing of the training set}

\begin{enumerate}
\item \textbf{Dataset reflections.}  
      Several parameters are bounded by physics—for example, the spin magnitude \(0 \le \chi \le 1\).  Posterior samples often accumulate at these limits, creating sharp density cliffs that density estimators struggle to model accurately.  When such an abrupt edge is detected, we reflect the samples across the boundary and down-weight the mirrored points with an exponential factor.  This procedure smooths the effective density while leaving the interior region of interest unchanged.

\item \textbf{Domain transforms.}  
      Density estimators operate in an unbounded latent space, each bounded coordinate \(x \in [a,b]\) is first linearly rescaled to the interval (0,1) and then mapped via the logit transform
\(z=\log\!\bigl[(x-a)/(b-a)\bigr]\).
The corresponding Jacobian is added to the probability estimation, ensuring that the overall density remains properly normalised.
\end{enumerate}

\subsubsection*{2. Normalizing Flow architecture}

In the transformed space we employ a masked autoregressive flow~\citep{papamakarios2018maskedautoregressiveflowdensity,kingma2017improvingvariationalinferenceinverse} with three coupling layers.
Each layer is parameterised by a rational–quadratic spline bijection
with 16 knots~\citep{durkan2019neuralsplineflows}.  
The base distribution is an isotropic standard normal.  
This configuration achieves high accuracy while keeping the model
complexity modest.

\subsubsection*{3. Normalizing Flow Training}

The flow is trained by minimising the negative log-likelihood
(equivalently, the Kullback–Leibler divergence) using the Adam optimiser~\citep{kingma2017adammethodstochasticoptimization}.  
$10\%$ of the data are reserved for validation to monitor
over-fitting; training stops automatically when the validation loss ceases to improve.

\section{Population Model}
\label{appx:pop_model}
To carry out the likelihood marginalization (see Sec.~\ref{sec:marg_like_subset}), we adopt the GWTC-3~\citep{gwtc-3_pop} maximum-likelihood population model, comprising separate descriptions for the binary black hole mass, spin, and redshift distributions. We detail each of these model components below:

\subsection{Mass Distribution Model}

We employ a \textsc{Power Law + Peak}~ \citep{Talbot_2018} mass distribution, described by:
\begin{equation}
\begin{aligned}
\pi(m_1 \mid \lambda_{\mathrm{mass}}) &= (1-\lambda_{\mathrm{peak}}) \mathcal{P}(m_1 \mid -\alpha, m_{\min}, m_{\max}) S(m_1 \mid m_{\min}, \delta_m) \\
&\quad+ \lambda_{\mathrm{peak}} \mathcal{G}(m_1 \mid \mu_m, \sigma_m) S(m_1 \mid m_{min}, \delta_m),
\end{aligned}
\end{equation}
where the terms are defined as follows:
\begin{itemize}
\item $\mathcal{P}(m_1 \mid -\alpha, m_{\min}, m_{\max})$ is a truncated power-law distribution characterized by a spectral index $-\alpha$, and mass limits $m_{\min}$, $m_{\max}$.
\item $\mathcal{G}(m_1 \mid \mu_m, \sigma_m)$ is a Gaussian distribution centered at $\mu_m$ with width $\sigma_m$.
\item $\lambda_{\mathrm{peak}}$ is the mixing fraction of mergers from the Gaussian peak.
\item $S(m_1 \mid m_{\min}, \delta_m)$ is a smoothing function tapering the distribution smoothly from zero to one over $[m_{\min}, m_{\min}+\delta_m]$:
\begin{equation}
S(m \mid m_{\min}, \delta_m) = \begin{cases}
0 & m < m_{\min}, \\
\left[ \exp(\frac{\delta_m}{m-m_{min}}+\frac{\delta_m}{(m-m_{min})-\delta_m}) +1 \right]^{-1} & m_{\min} \le m < m_{\min} + \delta_m, \\
1 & m \ge m_{\min} + \delta_m.
\end{cases}
\end{equation}
\end{itemize}

In addition, to accommodate events with masses exceeding the maximum mass $m_{max}$ of the primary mass distribution, we introduce an additional mixture component:
\begin{equation}
    \pi(m_1) = f_{\text{U}}\,U(m_1) + (1 - f_{\text{U}})\,\pi(m_1 \mid \lambda_{\mathrm{mass}}),
\end{equation}
where $U(m_1)$ is a uniform distribution spanning a sufficiently broad mass range, and we set the mixing fraction to $f_{\text{U}} = \frac{1}{200}$, which corresponds approximately to the inverse of the number of events analyzed.

The conditional mass-ratio distribution, $q = m_2 / m_1$, is modeled as a power-law:
\begin{equation}
\pi(q \mid \beta_q, m_1, m_{\min}) \propto q^{\beta_q}, \quad \text{for} \quad m_{\min} \le m_2 \le m_1.
\end{equation}

We fix the parameters of our mass model to the values presented in Table~\ref{tab:mass_model_parameters}.
\begin{table}[h!]
\centering
\begin{tabular}{ccc}
\hline
Parameter & Description & Value \\
\hline
$\alpha$ & Spectral index of primary mass distribution & 3.57 \\
$\beta_q$ & Spectral index of mass-ratio distribution & 0.63 \\
$m_{\min}$ & Minimum mass of power-law component & 5.30 $M_\odot$ \\
$m_{\max}$ & Maximum mass of power-law component & 88.69 $M_\odot$ \\
$\lambda_{\mathrm{peak}}$ & Gaussian component mixing fraction & 0.029 \\
$\mu_m$ & Mean of Gaussian peak & 34.60 $M_\odot$ \\
$\sigma_m$ & Width of Gaussian peak & 3.53 $M_\odot$ \\
$\delta_m$ & Mass tapering range at lower end & 4.34 $M_\odot$ \\
$f_{\text{U}}$ & Uniform distribution  mixing fraction & 0.005 \\
\hline
\end{tabular}
\caption{Adopted parameters for the \textsc{Power Law + Peak} mass model in this analysis.}
\label{tab:mass_model_parameters}
\end{table}

This explicit choice of parameter values ensures consistency with the best-fit population models presented in the GWTC-3 analysis.

\subsection{Spin population model}
We utilize the spin model from~\citep{Talbot_2018} introduced in GWTC-3 population analysis~\cite{gwtc-3_pop} following~\citep{PhysRevD.100.043012}. The dimensionless spin magnitudes $\chi_{1,2}$ follows a beta distribution,
\begin{equation}
\pi(\chi_{1,2}\mid\mu_\chi,\sigma_\chi) = \mathrm{Beta}(\mu_\chi,\sigma_\chi),
\end{equation}
where $\mu_\chi$ and $\sigma_\chi$ are shape parameters governing the mean and variance of the distribution. The spin magnitudes for the primary and secondary black holes, $\chi_1$ and $\chi_2$, are assumed to be independently and identically distributed.

The spin tilt angles $\cos\theta_i$, defined as the cosine of the angle between each component spin and the orbital angular momentum, are modelled as a mixture of two populations~\citep{PhysRevD.96.023012}:
\begin{equation}
\pi(\cos\theta_i\mid\zeta,\sigma_t) = \zeta G_t(\cos\theta_i\mid\sigma_t) + (1-\zeta) \mathcal{I}(\cos\theta_i),
\end{equation}
where $\mathcal{I}(\cos\theta_i)$ is an isotropic distribution, and $G_t(\cos\theta_i\mid\sigma_t)$ is a truncated Gaussian distribution peaked at maximal alignment $(\cos\theta_i = 1)$ with width $\sigma_t$. The mixing parameter $\zeta$ specifies the fraction of mergers originating from the aligned Gaussian component.

The parameters of this model, along with the values adopted in this analysis, are summarized in Table~\ref{tab:spin_params}.

\begin{table}[ht!]
\caption{Summary of spin model parameters and their adopted values.}
\label{tab:spin_params}
\centering
\begin{tabular}{lll}
\hline\hline
Parameter & Description & Value \\
\hline
$\mu_\chi$ & Mean of the beta distribution of spin magnitudes & 0.23 \\
$\sigma^2_\chi$ & Variance of the beta distribution of spin magnitudes & 0.033 \\
$\zeta$ & Mixing fraction of aligned-spin mergers & 0.97 \\
$\sigma_t$ & Width of truncated Gaussian in spin misalignment & 1.18 \\
\hline\hline
\end{tabular}
\end{table}
\subsection{Redshift evolution model}
\label{appx_subsec:redshift_evolution}
We adopt the commonly used \textsc{Power Law}~\citep{Fishbach_2018} redshift evolution model, which parameterizes the binary black hole merger rate density per comoving volume and source time as:
\begin{equation}
    \mathcal{R}(z) = \mathcal{R}_0(1 + z)^{\kappa},
\end{equation}
where \(\mathcal{R}_0\) denotes the merger rate density at \(z = 0\), and \(\kappa= 3.46\) is the redshift evolution slope parameter.
This implies that the redshift distribution scales with redshift as
\begin{equation}
    P(z) \propto \frac{dV_c}{dz}\frac{1}{1+z}(1 + z)^{\kappa},
\end{equation}
where ${dV_c}/{dz}$ is the differential comoving volume element and the factor $1/(1+z)$ compensates for cosmological time-dilation between the source frame and the detector frame.

\section{Implementation of effective-spin likelihood weights}
\label{appx:chi_eff_likelihood_weights}

In constructing likelihoods for the effective inspiral spin parameter $\chi_{\mathrm{eff}}$, one cannot impose a uniform prior distribution on $\chi_{\mathrm{eff}}$ while keeping \emph{all} the other spin population models the same. Therefore,  we marginalize over the entire parameter set except for the primary spin magnitude $\chi_1$. The parameter set considered is:
$$
\eta = (m_1, q, z, \chi_2, \cos\theta_1, \cos\theta_2),
$$
where the primary spin magnitude $\chi_1$ is determined uniquely from $\chi_{\mathrm{eff}}$ through the inverse transformation:
$$
\chi_1 = \frac{(1+q)\,\chi_{\mathrm{eff}} - q\,\chi_2\cos\theta_2}{\cos\theta_1}.
$$

The joint prior density $\pi(\eta, \chi_{\mathrm{eff}})$ can thus be expressed in two equivalent forms:
\begin{enumerate}
\item Conditioning explicitly on $\chi_{\mathrm{eff}}$:
\begin{align}
\label{eq:chi_eff_prior}
\pi(\eta,\chi_{\mathrm{eff}}) &= \pi(\chi_{\mathrm{eff}}\mid \chi_2,\cos\theta_1,\cos\theta_2,q)\pi(\eta) \nonumber \\
\pi(\eta) &= \frac{\pi(\eta,\chi_{\mathrm{eff}})}{\pi(\chi_{\mathrm{eff}}\mid \chi_2,\cos\theta_1,\cos\theta_2,q)}
\end{align}

\item Utilizing a change-of-variables formulation:
\begin{align}
\label{eq:chi_eff_jacobian}
\pi(\eta,\chi_{\mathrm{eff}}) = \pi(\eta,\chi_1)\left|\frac{\partial \chi_1}{\partial\chi_{\mathrm{eff}}}\right|.
\end{align}
\end{enumerate}

Following Eq.~\eqref{eq:marg_weights}, and using the relations above, the weights required to re-weight the posterior samples for a uniform prior in $\chi_{\mathrm{eff}}$ are given by:
\begin{align}
w_j &= \frac{\pi(\eta_j\mid \hat{\Lambda})}{\pi(\eta_j,\chi_{\mathrm{eff},j}\mid\text{\o})} \nonumber \\
&= \frac{\pi(\eta_j,\chi_{\mathrm{eff},j}\mid \hat{\Lambda})}{\pi(\chi_{\mathrm{eff},j}\mid \chi_{2,j},\cos\theta_{1,j},\cos\theta_{2,j},q_j,\hat{\Lambda})}\frac{1}{\pi(\eta_j,\chi_{\mathrm{eff},j}\mid\text{\o})} \nonumber \\
&=\frac{\pi(\eta_j,\chi_{1,j}\mid \hat{\Lambda})}{\pi(\eta_j,\chi_{1,j}\mid \text{\o})}
\frac{1}{\pi(\chi_{1,j} \mid \hat{\Lambda})}\left|\frac{\partial \chi_{1,j}}{\partial \chi_{\mathrm{eff},j}}\right|^{-1},
\label{eq:marg_weight}
\end{align}
where in the final step we used Eq.~\eqref{eq:chi_eff_jacobian}, and the Jacobian factor is given analytically by:

$$
\left|\frac{\partial \chi_1}{\partial \chi_{\mathrm{eff}}}\right| = \frac{1+q}{|\cos\theta_1|}.
$$

An equivalent weighting procedure can be formulated by reversing the conditioning in the prior. Instead of
\(\pi(\eta,\chi_{\mathrm{eff}})=\pi(\chi_{\mathrm{eff}}\mid \chi_2,\cos\theta_1,\cos\theta_2,q)\,\pi(\eta)\) (Eq.~\eqref{eq:chi_eff_prior}),
one may choose
\[
\pi(\eta,\chi_{\mathrm{eff}})=\pi(\chi_2,\cos\theta_1,\cos\theta_2,q \mid \chi_{\mathrm{eff}})\,\pi(\chi_{\mathrm{eff}}).
\]
Under this parameterization, the resulting sample–weight expression (Eq.~\eqref{eq:marg_weights}) would contain
\(\pi(\chi_{\mathrm{eff}})\) explicitly in the denominator, which would typically require numerical evaluation--for example, via a kernel density estimator. By contrast, in Eq.~\eqref{eq:marg_weights} all terms are analytic, enabling an efficient implementation without significant computational overhead. 
The two approaches are equivalent and should converge to the same result.

The implementation for
\(\chi_{\mathrm{p}}\) is analogous.

\section{Optimization Algorithm}
\label{app:pistroke_algorithm}

\begin{figure}[ht]
  \centering
  \includegraphics[width=0.99\textwidth]{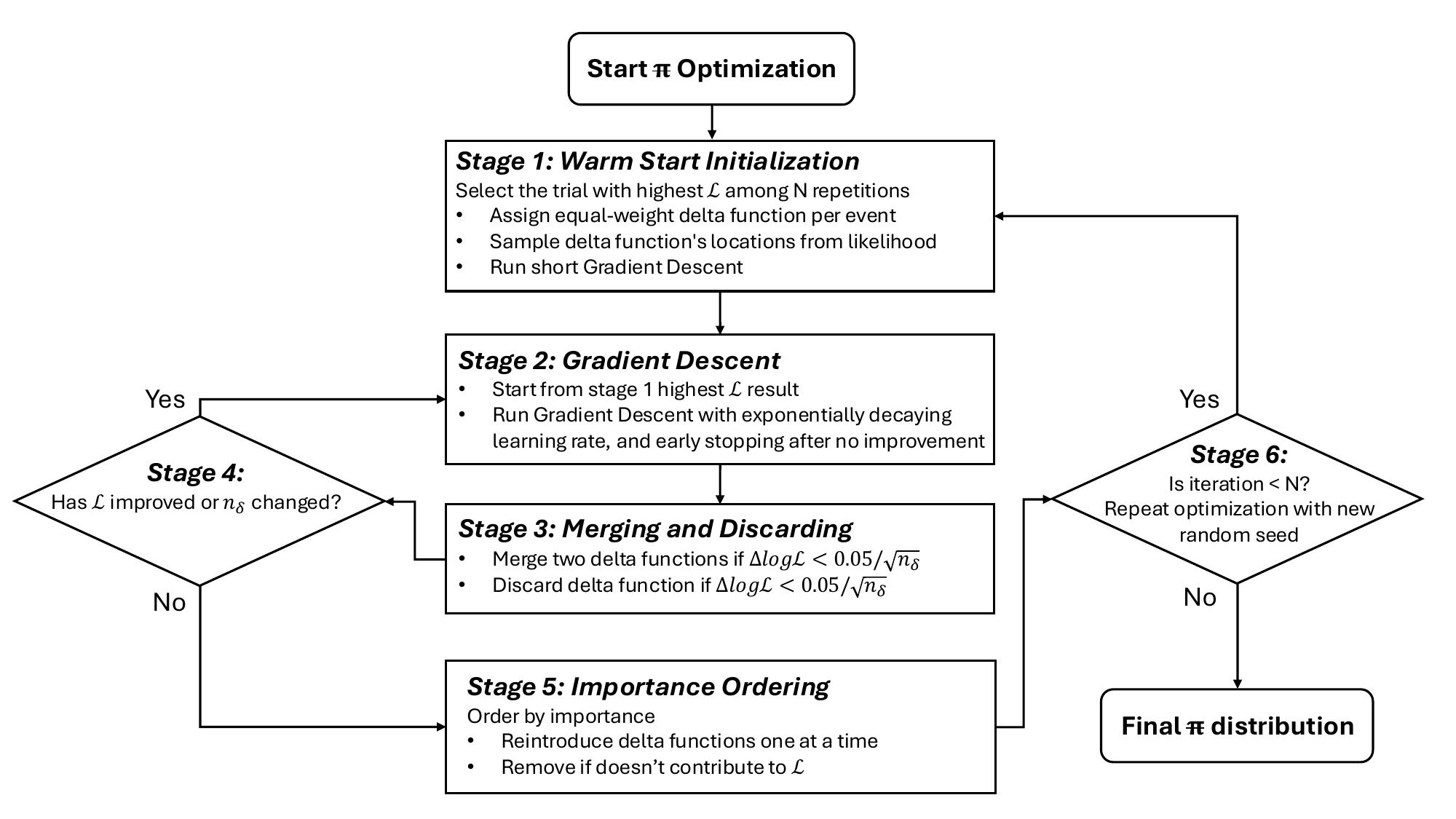}
  \caption{Flowchart of the $\pistroke$-optimization procedure. Rounded rectangles mark the entry and terminal state. Plain rectangles (stages 1, 2, 3, and 5) indicate algorithmic steps: warm-start initialization, gradient descent with decay/early stopping, merging/discarding of delta functions, and importance reordering/reintroduction. Diamonds represent decision points (stages 4 and 6). Solid arrows show the direction of execution, and ‘Yes/No’ labels indicate the branch taken. $\mathcal{L}$ denotes the population likelihood, $\Delta\!\log\mathcal{L}$ denotes the change in log-likelihood, $n_{\delta}$ the number of delta functions, and N the maximum number of random-seed restarts.}
  \label{fig:pistroke_flowchart}
\end{figure}

Figure~\ref{fig:pistroke_flowchart} summarizes the $\pistroke$-optimization procedure described in Section~\ref{sec:optimisation} for constructing the maximum-likelihood. Rectangles denote processing stages, diamonds denote decision tests, rounded rectangles denote the start/end states, and arrows indicate control flow, with symbols defined as follows: $\mathcal{L}$ is the population likelihood, $\Delta\!\log\mathcal{L}$ is the log-likelihood change, $n_{\delta}$ the number of delta functions, and N the maximum number of random-seed restarts.

\section{Additional $\pistroke$ analysis}
\label{appx:additional__pi_stroke}
\begin{figure}[ht!]
  \centering
  \begin{minipage}{.48\linewidth}
    \centering
    \includegraphics[width=\linewidth]{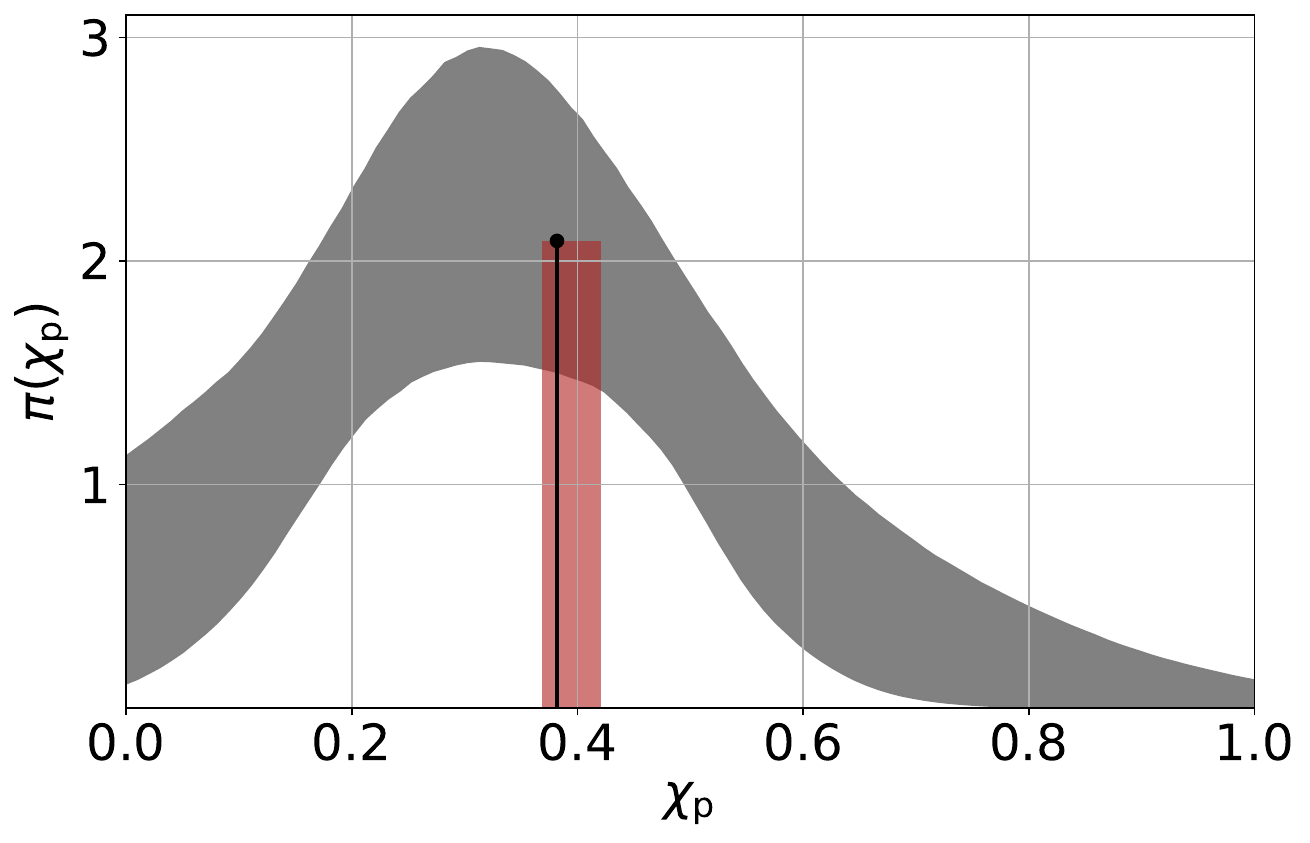}
  \end{minipage}
  \hfill
  \begin{minipage}{.48\linewidth}
    \centering
    \includegraphics[width=\linewidth]{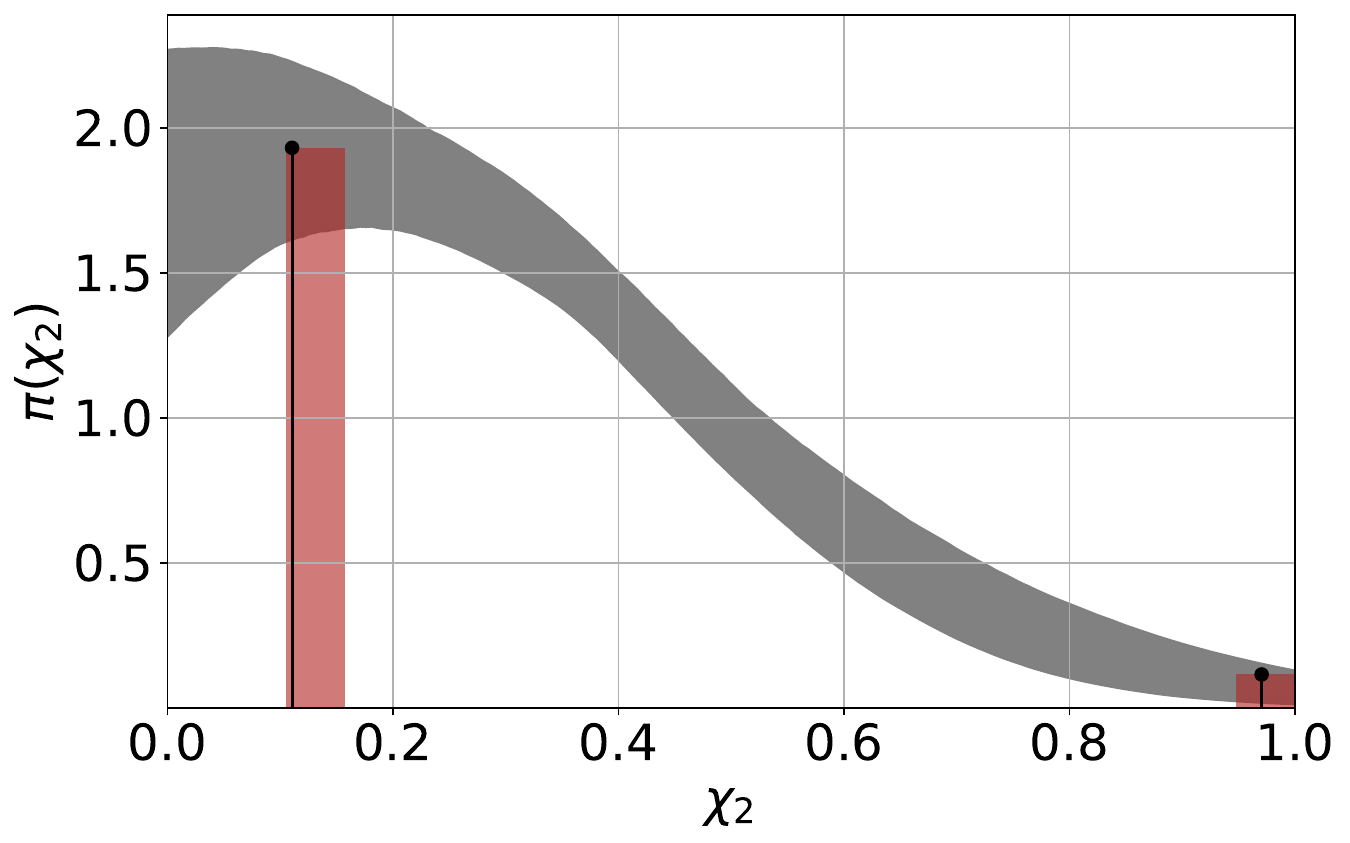}
  \end{minipage}
  \caption{One-dimensional $\pistroke$ distribution for \(\chi_{\mathrm{p}}\) (left) and \(\chi_2\) (right), both consistent with GWTC-4 parametric model. Plotting conventions match those of Fig.~\ref{fig:1d_m1}}
    \label{fig:1d_chi_p_chi2}
\end{figure}

For completeness, we present additional $\pistroke$ distributions for \(\chi_{\mathrm{p}}\), and $\chi_2$ in Fig.~\ref{fig:1d_chi_p_chi2}.

\section{Two-dimensional marginals}
\label{appx:compare_1D_to_2D}

Here we compare the two-dimensional $\pistroke$ marginal distributions with the corresponding one-dimensional $\pistroke$ reconstructions. 

In general, perfect agreement is not expected: when the two parameters are independent, the two-dimensional marginals should closely match the one-dimensional results, when there is correlation or other statistical dependence, noticeable deviations are expected.

In each figure, black markers and vertical lines show the delta functions from the two-dimensional analysis after marginalizing over the second parameter; their heights are proportional to the associated weights. The red bars give a histogram of these $\pistroke$ samples. Blue markers and lines denote the delta functions from the one-dimensional analysis.

\begin{figure}[h]
  \centering
  \includegraphics[width=\textwidth]{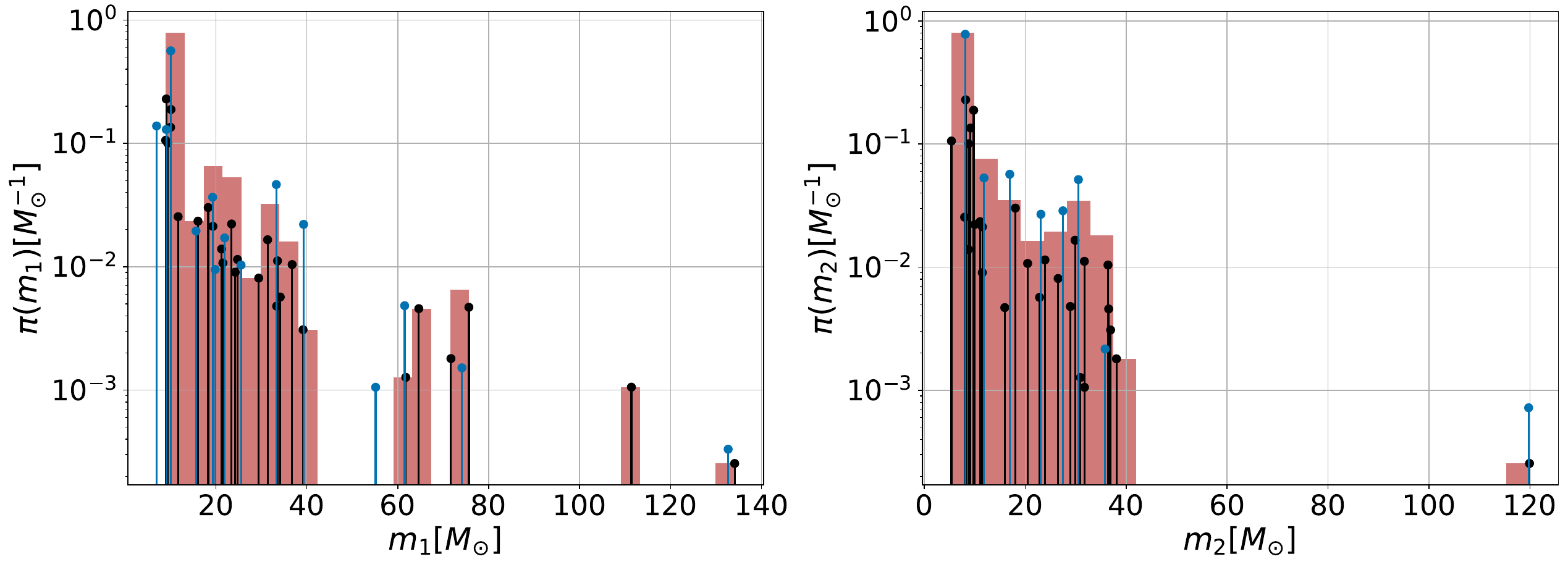}
  \caption{The one-dimensional (blue) and two-dimensional (black and red) marginal $\pistroke$ distributions for $m_1$ (left) and $m_2$ (right).}
  \label{fig:marg_m1_m2}
\end{figure}

\begin{figure}[h]
  \centering
  \includegraphics[width=\textwidth]{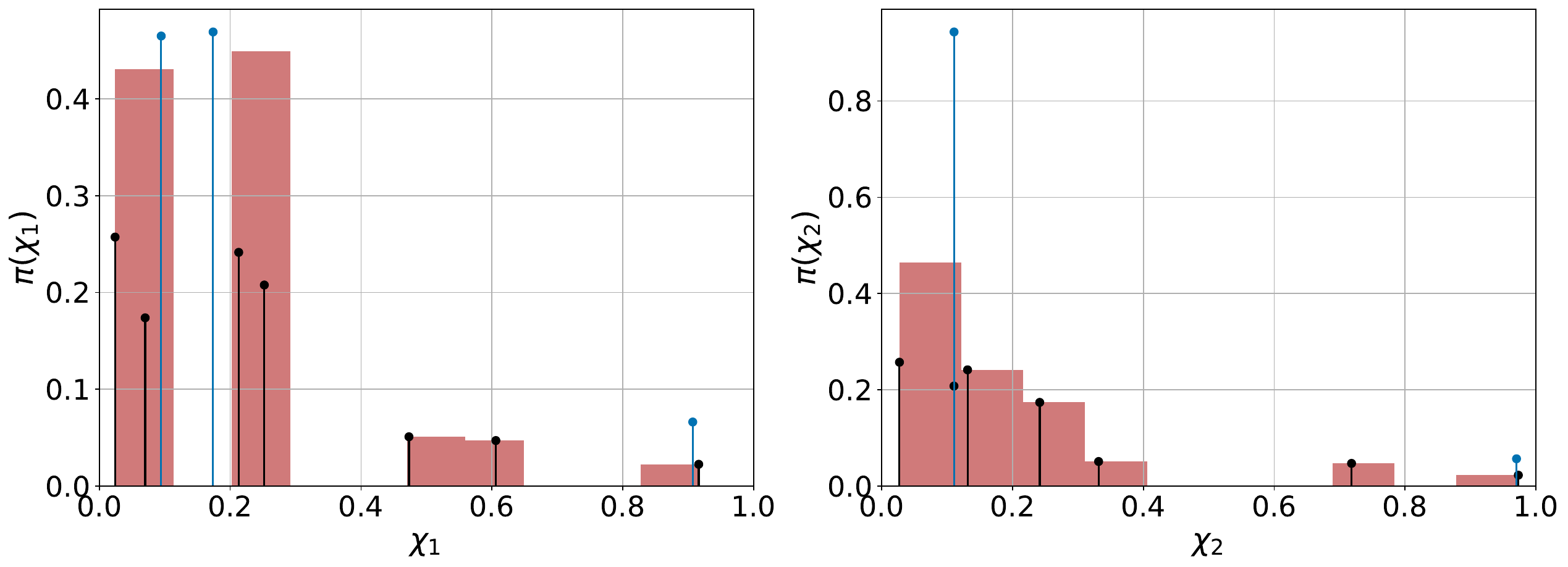}
   \caption{The one-dimensional (blue) and two-dimensional (black and red) marginal $\pistroke$ distributions for $\chi_1$ (left) and $\chi_2$ (right).}
  \label{fig:marg_chi1_chi2}
\end{figure}

\begin{figure}[h]
  \centering
  \includegraphics[width=\textwidth]{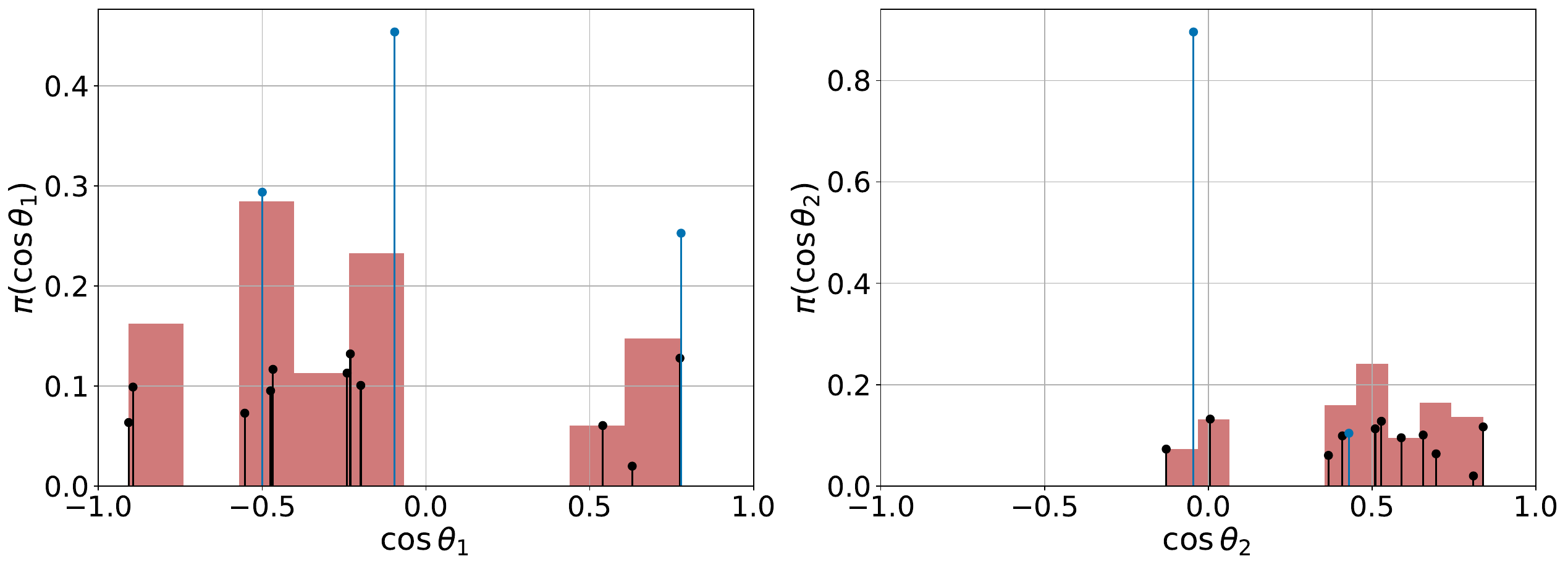}
    \caption{The one-dimensional (blue) and two-dimensional (black and red) marginal $\pistroke$ distributions for $\cos\theta_1$ (left) and $\cos\theta_2$ (right).}
  \label{fig:marg_cos1_cos2}
\end{figure}

Figure~\ref{fig:marg_m1_m2} illustrates the (\(m_1\), \(m_2\)) case; corresponding figures for (\(\chi_1\) , \(\chi_2\)) and (\(\cos\theta_1\), \(\cos\theta_2\)) are given in Figs.~\ref{fig:marg_chi1_chi2} and \ref{fig:marg_cos1_cos2}.

For both \((\chi_1,\chi_2)\) and \((\cos\theta_1,\cos\theta_2)\), we find clear—though not dramatic—differences between the one-dimensional results and the marginals of the two-dimensional analysis, most notably for the secondary component (\(\chi_2\) and \(\cos\theta_2\)). This behavior hints at statistical dependence between the components.

\bibliographystyle{aasjournal}
\bibliography{ref}

\end{document}